\documentclass{emulateapj}
\usepackage{apjfonts}
\usepackage{graphicx}
\usepackage{wrapfig}
\usepackage{amsmath}

\def\ltsim{\raise 2pt \hbox {$<$} \kern-1.1em \lower 4pt \hbox {$\sim$}}
\def\gtsim{\raise 2pt \hbox {$>$} \kern-1.1em \lower 4pt \hbox {$\sim$}}

\begin{document}

\shorttitle{Large-scale shock and cool filaments in Hydra A} 
\shortauthors{Gitti et al.}

\title{A Chandra study of the large-scale shock and cool
  filaments in Hydra A: Evidence for substantial gas dredge-up by the
  central outburst }

\author{Myriam Gitti\altaffilmark{1,2}, 
Paul E. J. Nulsen\altaffilmark{1}, 
Laurence P. David\altaffilmark{1}, 
Brian R. McNamara\altaffilmark{1,3,4}, 
and Michael W. Wise\altaffilmark{5}}

\altaffiltext{1}{Harvard-Smithsonian Center for Astrophysics, 60
  Garden Street, Cambridge, MA 02138 - USA} \altaffiltext{2}{
  Astronomical Observatory of Bologna - INAF, via Ranzani 1, I-40127
  Bologna - Italy} \altaffiltext{3}{ Dept. of Physics \& Astronomy,
  University of Waterloo, 200 University Avenue West, Waterloo,
  Ontario - Canada N2L 2G1} \altaffiltext{4}{Perimeter Institute for
  Theoretical Physics, Waterloo, Canada} \altaffiltext{5}{ASTRON
  Netherlands Institute for Radio Astronomy, P.O. Box 2, 7990 AA
  Dwingeloo - The Netherlands }

\begin{abstract}
  We present the results of a Chandra study of the Hydra A galaxy
  cluster, where a powerful AGN outburst created a large-scale cocoon
  shock.  We investigated possible azimuthal variations in shock
  strength and shape, finding indications for a weak shock with a Mach
  number in the range $\sim$1.2-1.3. We measured the temperature
  change across the shock front. However, the detection of a
  temperature rise in the regions immediately inside of the front is
  complicated by the underlying temperature profile of the cluster
  atmosphere. We measured the global temperature profile of the
  cluster up to 700 kpc, which represents the farthest measurement
  obtained with Chandra for this cluster. A ``plateau'' in the
  temperature profile in the range $\sim$70-150 kpc indicates the
  presence of cool gas, which is likely the result of uplift of
  material by the AGN outburst. After masking the cool filaments
  visible in the hardness ratio map, the plateau disappears and the
  temperature profile recovers a typical shape with a peak around 190
  kpc, just inside the shock front. However, it is unlikely that such
  a temperature feature is produced by the shock as it is consistent
  with the general shape of the temperature profiles observed for
  relaxed galaxy clusters. We studied the spectral properties of the
  cool filaments finding evidence that $\sim$10$^{11} M_{\odot}$ of
  low-entropy material has been dredged up by the rising lobes from
  the central 30 kpc to the observed current position of 75-150 kpc.
  The energy required to lift the cool gas is \gtsim $2.2 \times
  10^{60}$ erg, which is comparable to the work required to inflate
  the cavities and is $\sim$25\% of the total energy of the
  large-scale shock.  Our results show that the AGN feedback in Hydra
  A is acting not only by directly heating the gas, but also by
  removing a substantial amount of potential fuel for the SMBH.
\end{abstract}


\maketitle


\section{Introduction}

Observational and theoretical evidence has been growing in the past
decade in favor of the existence in galaxy clusters of a feedback
mechanism that prevents cool cores from establishing ``cooling flows''
at the rates predicted by earlier X-ray observations (e.g., Peterson
\& Fabian 2006 for a review).  The dominant cD galaxies, which are
present at the cluster center in all cool core clusters, host the most
massive black holes in the local Universe and usually show nuclear
activity. They accordingly provide a natural feedback mechanism for
the regulation of the cooling process.  Feedback is also required to
suppress the overproduction of massive galaxies predicted by
dark-matter-only simulations and to break the self-similarity of
clusters (e.g., Benson et al. 2003).  The nature of this feedback,
vital to our understanding of galaxy and structure evolution, is one
of the most important unresolved questions in extragalactic astronomy.
Based on the detection of cavities and AGN-driven shocks, the primary
source of feedback in cluster has been identified as radio galaxies
acting through outbursts and accompanying energy injection, likely
intermittent, from the central AGN (e.g., for a review, McNamara \&
Nulsen 2007 and references therein).  In galaxy clusters, where
cooling rates should be highest, the current generation of X-ray
observatories {\it Chandra} and {\it XMM-Newton} have shown that there
is not a significant amount of gas cooling below about one third of
its virial temperature (Peterson et al. 2003; Kaastra et
al. 2004). Images from these telescopes also reveal highly disturbed
structures in the cores of many clusters, including shocks, cavities
and sharp density discontinuities. At radio wavelengths, it is clear
that AGN jets are the cause of many of these disturbances.  The
incidence and variety of bubbles, cavities, shocks, and ripples
observed both in the radio and in X-rays in galaxy clusters provides
direct evidence of the widespread presence of AGN-driven phenomena
(e.g., Fabian et al. 2003 for a discussion of the Perseus cluster
properties in terms of AGN-generated viscously-damped sound waves;
Nulsen et al. 2005 for shock heating in Hydra A; Gitti et al. 2007 for
a study of giant cavities created by the most powerful AGN outburst
currently known; B\^irzan et al. 2008 and Diehl et al. 2008 for a
survey of cavities and the implied cavity heating rates). Such AGN
feedback has a wide range of impacts, from the formation of galaxies,
through to the explanation of the observed $M_{\rm bh}$-$\sigma$
relation (which indicates a causal connection or feedback mechanism
between the formation of bulges and their central black holes, e.g.,
Magorrian et al. 1998), to the regulation of cool cores. In most
cases, the energy introduced by the AGN is more than sufficient to
counteract putative cooling flows (B\^{\i}rzan et al. 2004, 2008;
Rafferty et al. 2006).
However, the details of how the feedback loop operates are still
unknown.  Only by studying striking examples of interaction between
the central radio galaxy and the ICM can we understand why cooling and
star formation still proceeds at a reduced rate, and this is likely to
reveal the coupling between AGN feedback and the ICM.


\begin{deluxetable}{ccccc}
\tabletypesize{\scriptsize}
\tablewidth{0pt}
\tablecaption{Shock properties
\label{shock_SB.tab}
}
\tablehead{
Sector  & Shock radius & Radius variation & Density jump & Mach number   \\
$\theta_1$-$\theta_2$ $(^{\circ})$ & $R_{\bar{\theta}}$ $(^{\prime\prime})$ & $R_{\theta_1}$-$R_{\theta_2}$ $(^{\prime\prime})$ & & $\mathcal{M}$ 
}
\startdata
0-30 & 208 & 201-222 & 1.33 & 1.22  
\\[+1mm]
30-60 & 257 & 222-303 & 1.45 & 1.30 
\\[+1mm]
60-90 & 332 & 314-349 & 1.33 & 1.23 
\\[+1mm]
90-120 & 371 & 364-342  &1.38 & 1.26 
\\[+1mm]
120-150 & 301 & 342-271 & 1.46 & 1.32 
\\[+1mm]
150-180 & 260 & 264-265 & 1.40 & 1.27 
\\[+1mm]
180-270 & 285 & 260-253 & 1.15 & 1.10  
\\[+1mm]
270-360 & 205 & 233-202 &1.35 & 1.23 
\\[-2mm]
\enddata
\tablecomments{ Results of the fits of a broken power-law density
  model to the surface brightness profiles extracted along different
  sectors. The sector aperture (from P.A. $\theta_1$ to $\theta_2$) is
  indicated in the first column.  The shock radius at the mid-angle of
  the sector ($R_{\bar{\theta}}$) is indicated in the second column,
  whereas the shock radii at the starting ($R_{\theta_1}$) and ending
  ($R_{\theta_1}$) angles are indicated in the third column.  Such
  radial distances are measured from the cluster center.  The
  corresponding density jump and Mach number ($\mathcal{M}$) are shown
  in the forth and fifth columns, respectively. Best-fit statistical
  errors are on average $\sim$5\%.
}
\end{deluxetable}


The galaxy cluster Hydra A has a well-known, large-scale system of
X-ray cavities embedded in a ``cocoon'' shock surrounding the central,
powerful radio source (McNamara et al. 2000, Nulsen et al. 2005). It
is considered one of the prototypes of cool core clusters with
cavities, which has served as an early test of the AGN feedback
paradigm, and it has been extensively studied both in the radio and
X-rays (Taylor et al. 1990; McNamara et al. 2000; David et al. 2001;
Nulsen et al. 2002, 2005; Lane et al. 2004; Wise et al. 2007;
Simionescu et al.  2009a, 2009b, Kirkpatrick et al. 2009).  By
analyzing the archival $\sim$200 ks {\it Chandra} exposure in this
paper we study the azimuthal properties of the large-scale shock and
attempt to measure the temperature jump associated with the shock in
different directions. We also perform a detailed spectral analysis of
the cool X-ray filaments extending out to 150 kpc, finding evidence
for extensive mass dredge-up from the central 30 kpc, which affects
the global temperature profile of the cluster.

With $H_0 = 70 \mbox{ km s}^{-1} \mbox{ Mpc}^{-1}$, and
$\Omega_M=1-\Omega_{\Lambda}=0.3$, the luminosity distance to Hydra A
(z=0.0538) is 240 Mpc and 1 arcsec corresponds to 1.05 kpc in the rest
frame of the cluster.


\section{Chandra observations and data reduction}


\begin{figure}
\centerline{
\includegraphics[width=8.5cm]{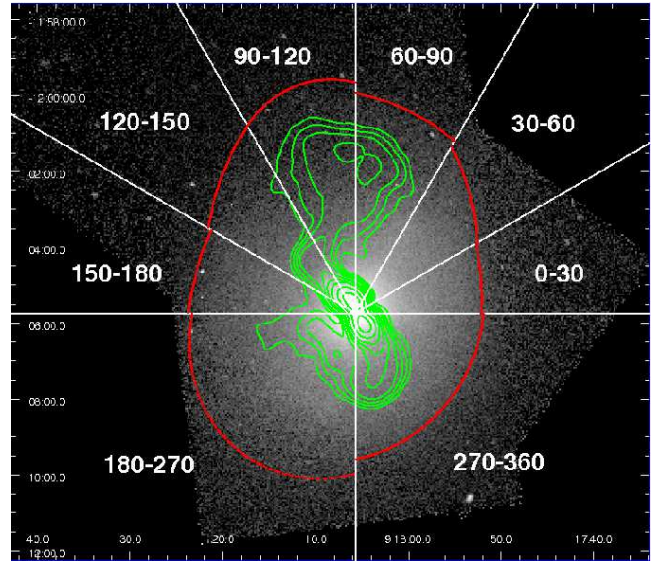}
}
\caption{\small \label{shock_image.fig} Residual map of the beta model
  subtracted surface brightness image from Nulsen et al. (2005).  The
  green contours outline the 330 MHz radio emission from Lane et
  al. (2004).  Overlaid in white are the sectors considered to
  investigate the azimuthal variations in shock properties by means of
  the surface brightness profile.  In each sector, the red curve
  indicates the position of the shock front determined by fitting a
  broken power-law density model to the surface brightness profile
  (see Fig. \ref{shock_SB.fig} and Sect. \ref{azimuthal.sec} for
  details). In the sector 180-270, where the surface brightness
    discontinuity is least evident, we have indicated the shock front
    with a dashed curve.}
\end{figure}


Hydra A has been imaged four times by {\it Chandra} ACIS for a total
exposure of 240 ks. The two shorter exposures (ObsIDs 575 and 576,
which collectively comprise only $\sim$17\% of the total exposure)
were taken early in the {\it Chandra} mission when the ACIS detector
was operated at the higher focal plane temperature of $-110
^{\circ}$C.  Due to the higher quality calibration at $-120
^{\circ}$C, the spectral analysis presented below was performed on the
two longer, more recent exposures: $\sim$97 ks collected on 2004
January 13 (ObsID 4969) and $\sim$99 ks collected on 2004 October 22
(ObsID 4970) with ACIS-S.  We use data from the S3 and S2 CCDs to
study the central part of the cluster emission where the cavity system
and the radio source are located, and from the I2 and I3 CCDs to
measure the temperature and surface brightness in the cluster
outskirts.  Each dataset was individually reprocessed with CIAO
version 4.1 using CALDB 4.1.0 and corrected for known time-dependent
gain and charge transfer inefficiency problems following techniques
similar to those described in the {\it Chandra} analysis
threads\footnote{http://cxc.harvard.edu/ciao/threads/index.html}.
Screening of the event files was also applied to filter out strong
background flares.  Blank-sky background files, filtered in the same
manner as in the Hydra A image and normalized to the count rate of the
source image in the 10-12 keV band, were used for background
subtraction.  We identified and removed the point sources using the
CIAO task {\ttfamily WAVDETECT}, with the detection threshold set to
the default value of $10^{-6}$.  The final, combined exposure time for
the two dataset is 174.2 ks.


\section{The Large-Scale Shock}

The observed feature interpreted as a shock front in the X-ray surface
brightness of Hydra A surrounds the low-frequency radio lobes, and the
correspondence between their shapes supports the interpretation of a
cocoon shock of the radio source.  Nulsen et al. (2005) estimated an
age of the outburst of $t_s = 1.4 \times 10^8$ yr and a total energy
of $E_s = 9 \times 10^{60}$ erg.  The shock front is clearly
aspherical with a complicated 3-D shape affected by projection
effetcs. In particular, the Hydra A radio source and cavity system is
inclined at about 40$^{\circ}$ to the plane of the sky, with the
northern side lying closer to us (Taylor 1996, Lane et al. 2004, Wise
et al. 2007).  The cavity and shock geometry as well as the outburst
history and energetics have been studied in detail in Wise et
al. (2007) and Simionescu et al. (2009b).


\subsection{Azimuthal Variations in Shock Strength}
\label{azimuthal.sec}


\begin{figure*}
\centerline{
\includegraphics[width=5.5cm, angle=-90]{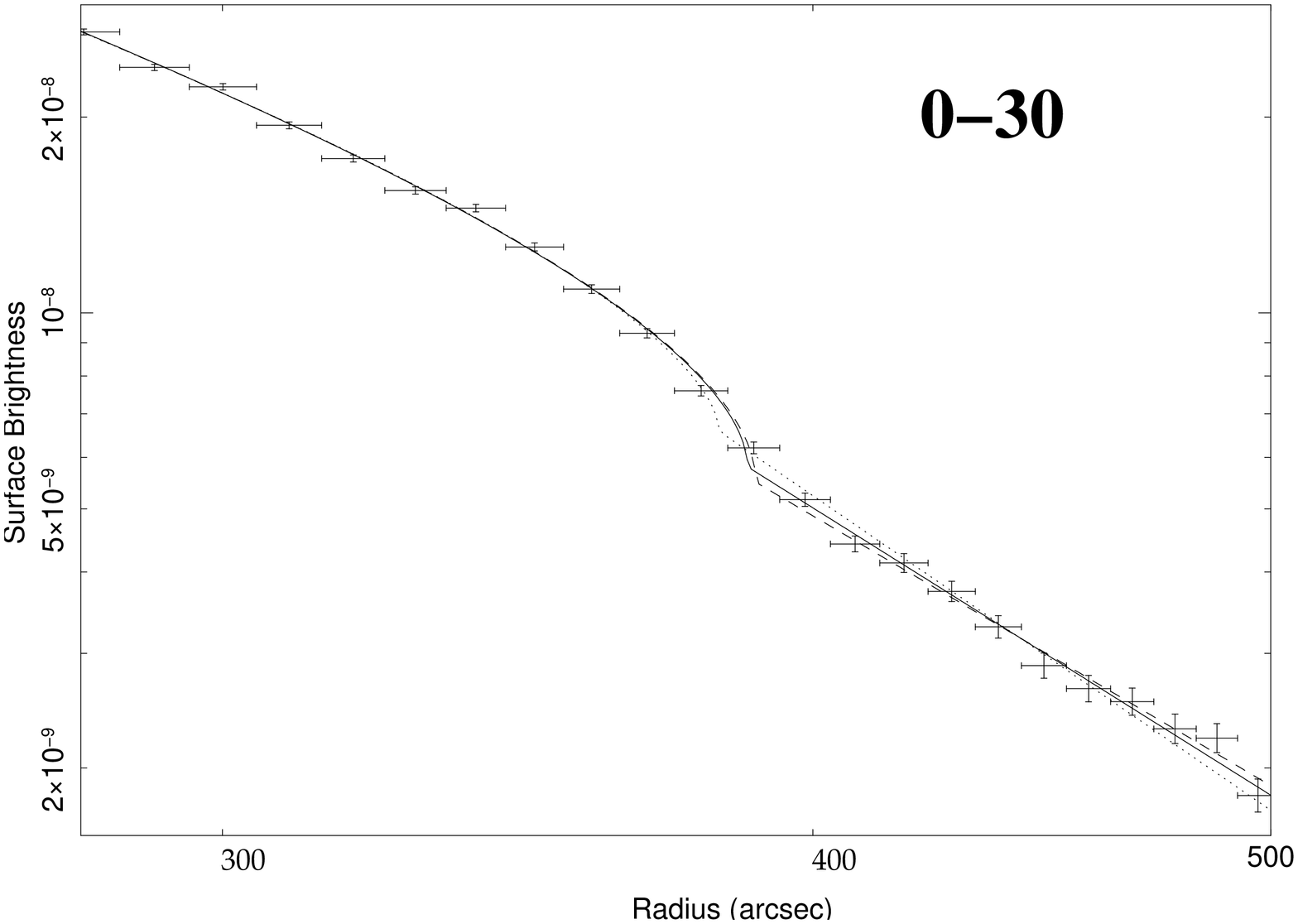}
\includegraphics[width=5.5cm, angle=-90]{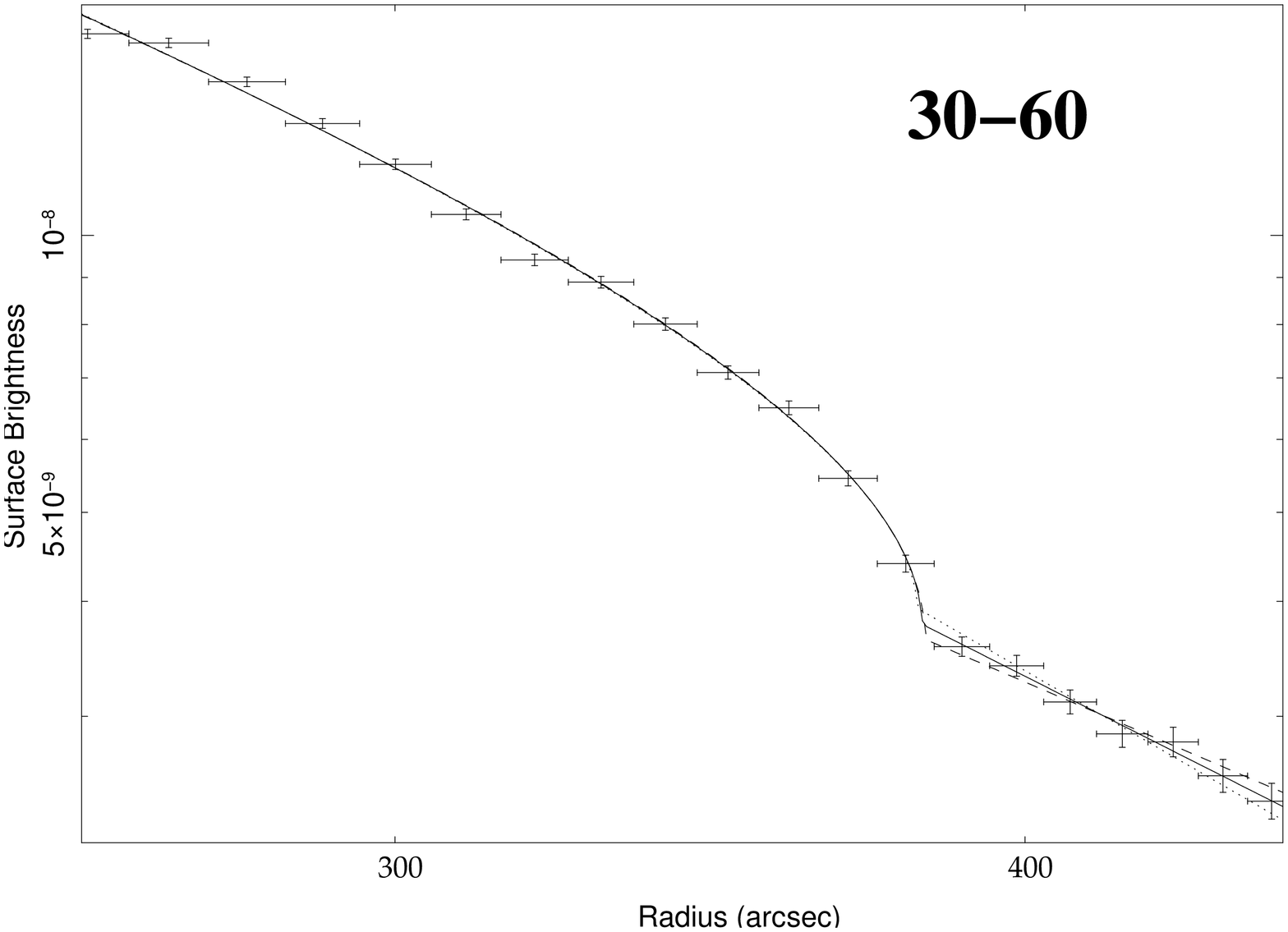}
}
\centerline{
\includegraphics[width=5.5cm, angle=-90]{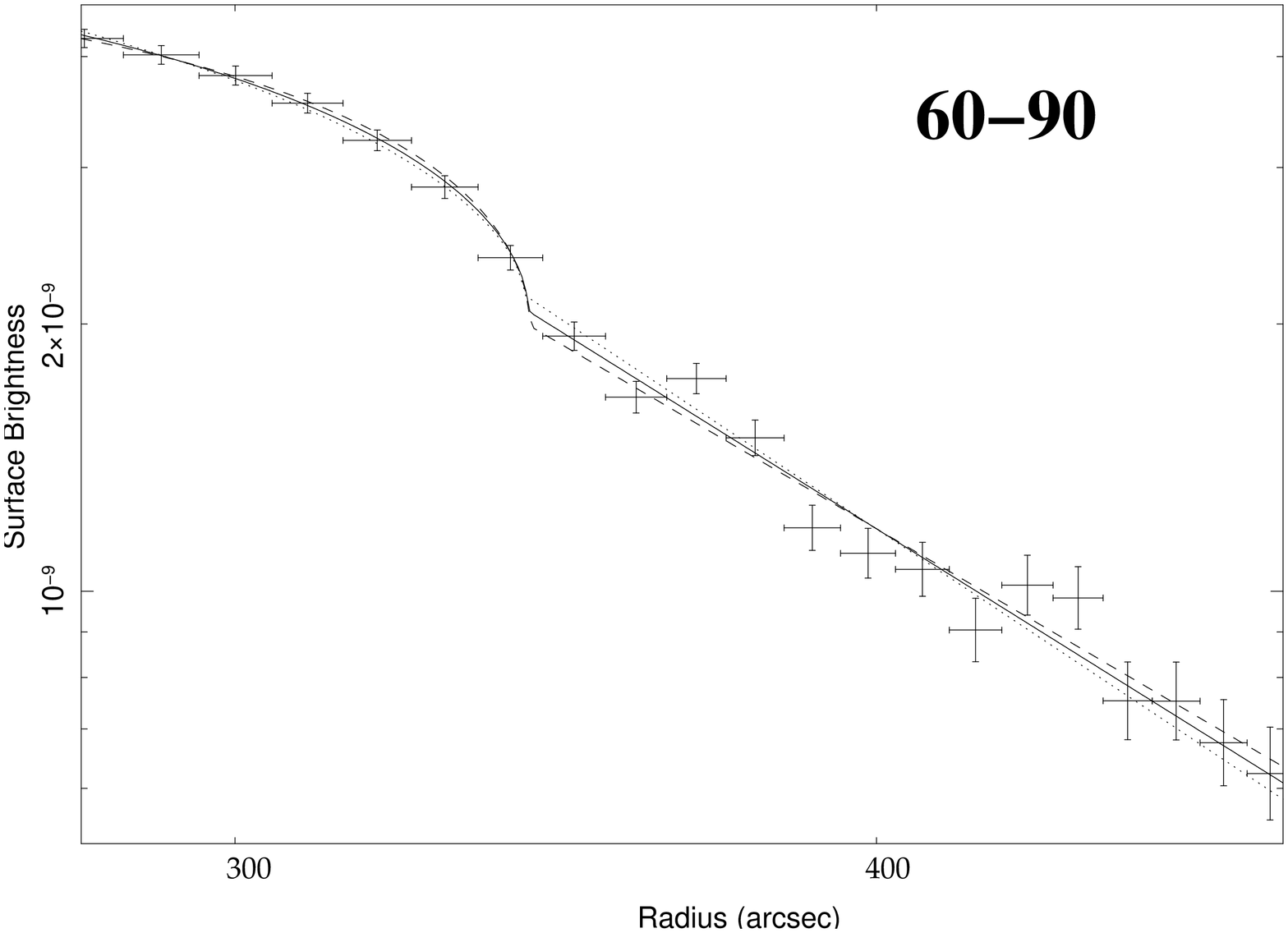}
\includegraphics[width=5.5cm, angle=-90]{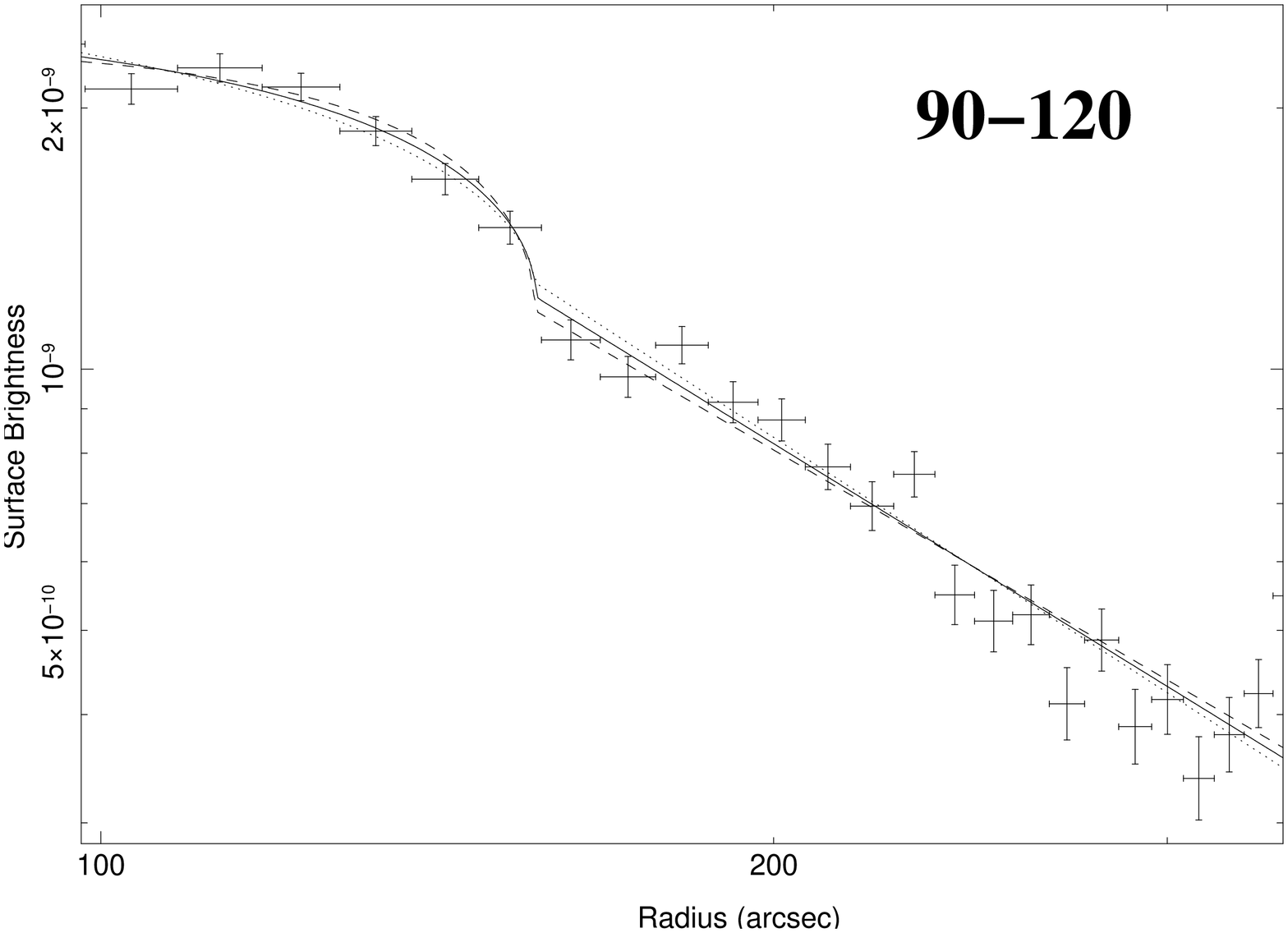}
}
\centerline{
\includegraphics[width=5.5cm, angle=-90]{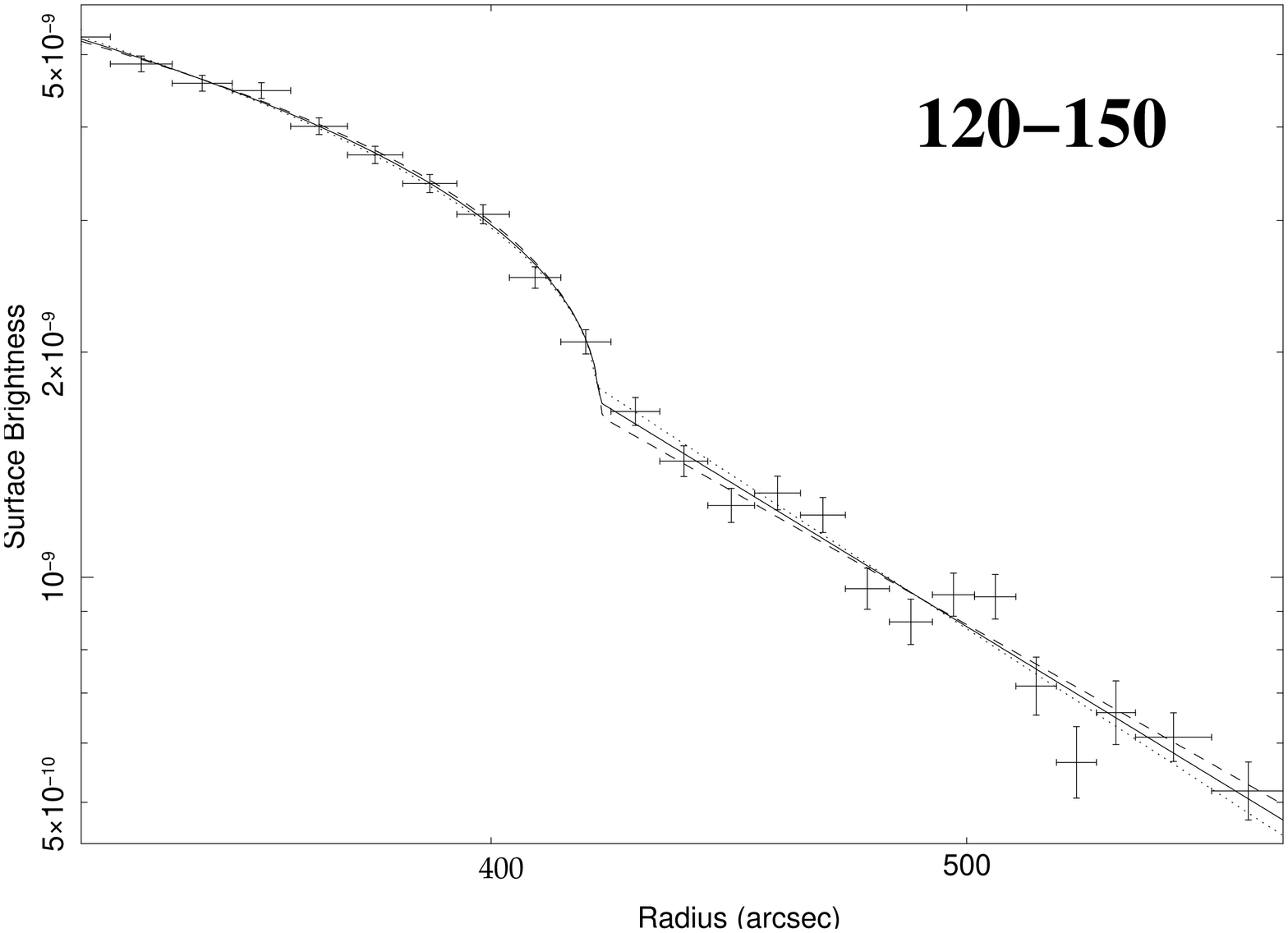}
\includegraphics[width=5.5cm, angle=-90]{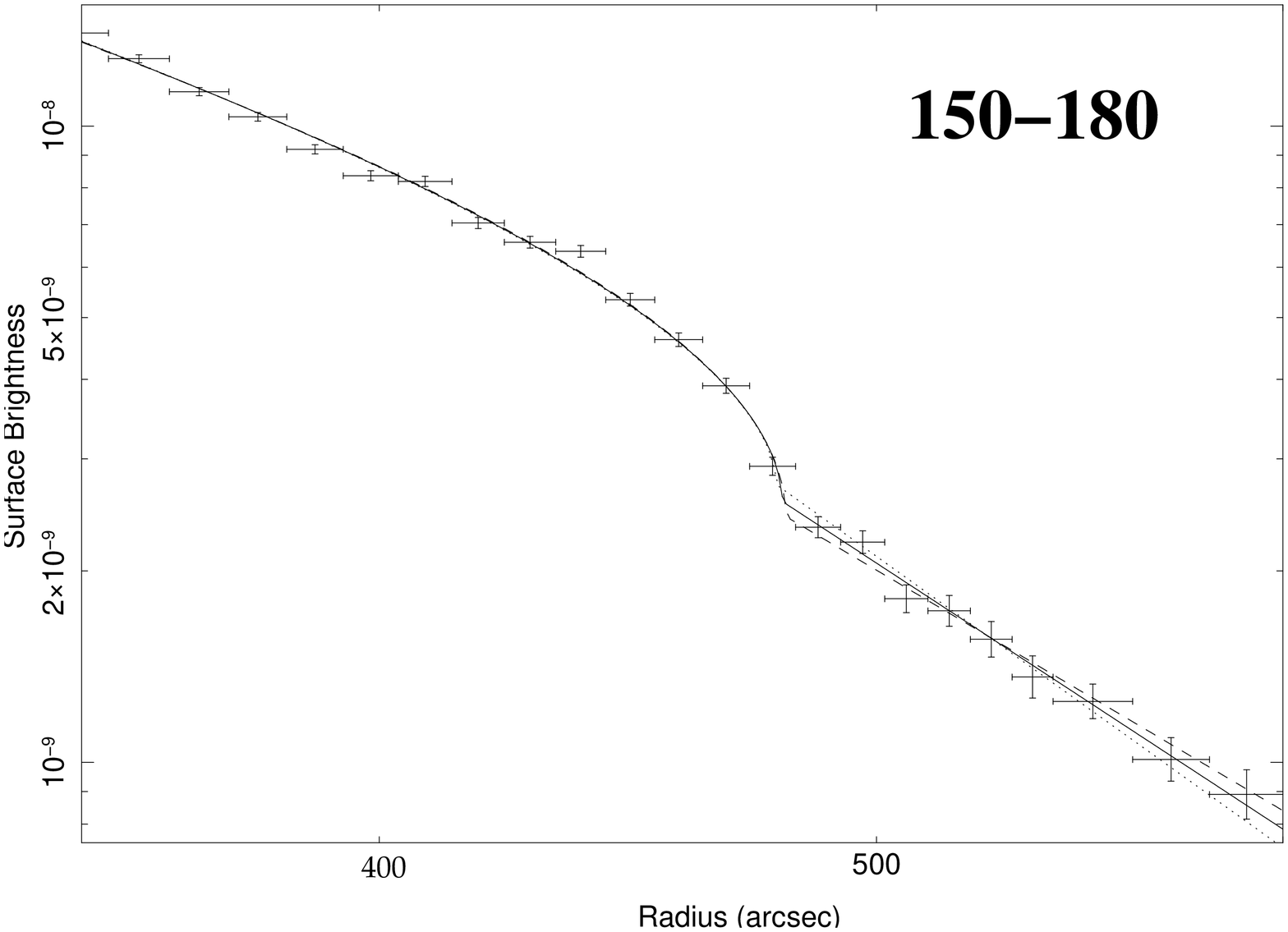}
}
\centerline{
\includegraphics[width=5.5cm, angle=-90]{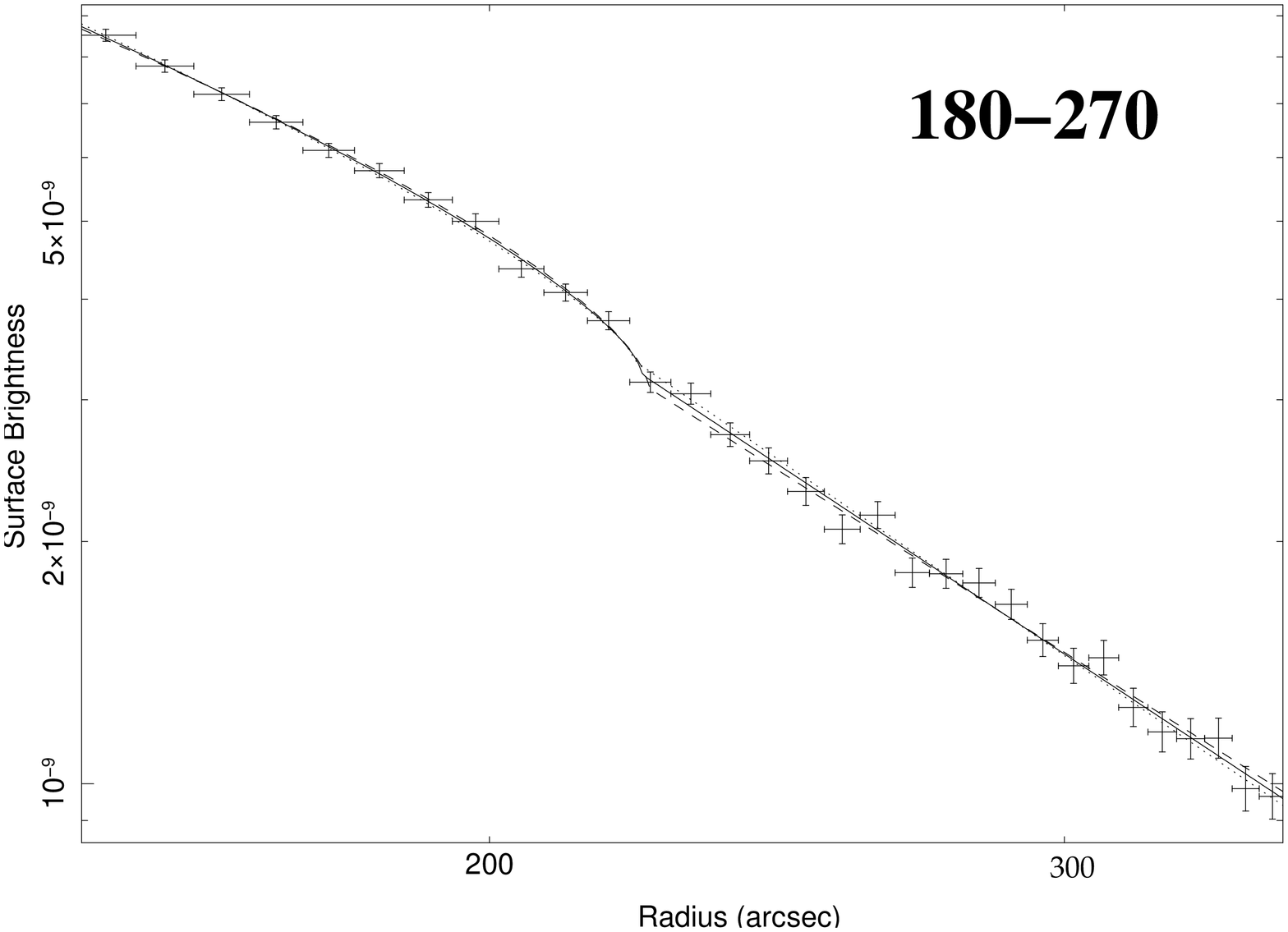}
\includegraphics[width=5.5cm, angle=-90]{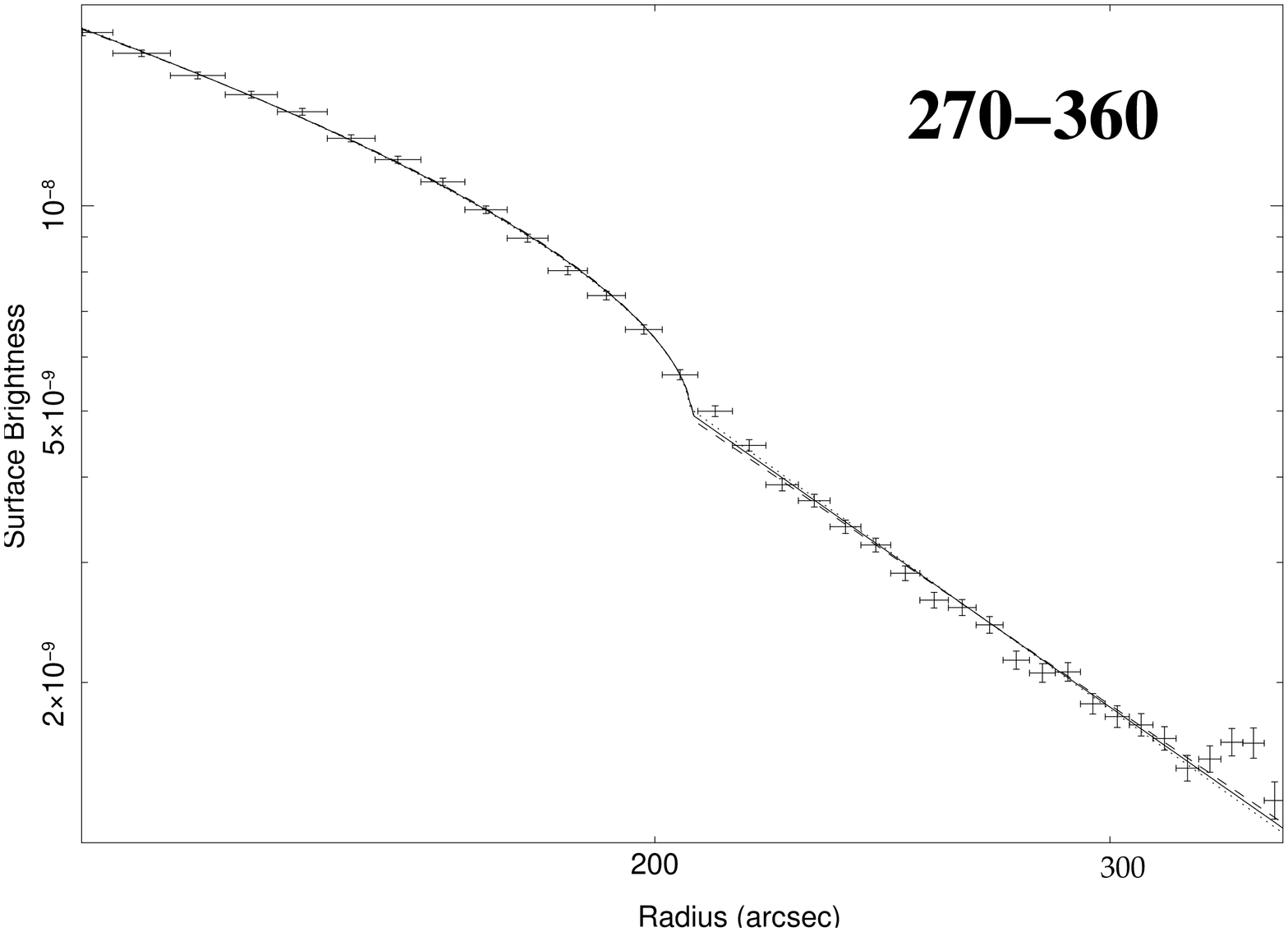}
}
\caption{\scriptsize \label{shock_SB.fig} Background-subtracted,
  exposure-corrected {\it Chandra} surface brightness profiles
  extracted along the sectors outlined in
  Fig. \ref{shock_image.fig}. The surface brightness is in units of
  counts cm$^{-2}$ s$^{-1}$, with errors at 1$\sigma$. The radial axis
  shows the distance from the center of curvature of the annuli used
  to extract the surface brightness profile. Radial error bars show
  the bin sizes. The smooth curves show fits of the broken power-law
  density model which give the density jumps and corresponding Mach
  numbers reported in Table \ref{shock_SB.tab}.  }
\end{figure*}


We have investigated the azimuthal variations in shock strength and
shape by studying surface brightness profiles in different
directions. In particular, we have divided the northern cluster
semicircle into six sectors of 30 degrees each, and the southern
cluster semicircle, where the shock is less evident in the image, into
two quadrants of 90 degrees each.  Starting from the West with
position angle (P.A.) = $0^{\circ}$ and counting counterclockwise, the
sectors are labeled as: 0-30, 30-60, 60-90, 90-120, 120-150, 150-180,
180-270, 270-360 (see Fig. \ref{shock_image.fig}).  We extracted the
background-subtracted, exposure-corrected surface brightness profile
along each sector in the energy range 0.5-2.0 keV.  The center of the
annular regions used to extract each profile was chosen in order to
best match the curvature of the radial bins with the shape of the
shock front as seen in the image (Fig. \ref{shock_image.fig}).  We
then performed fits of a broken power-law density model to each
surface brightness profile. This analysis assumes that the radius of
curvature in the plane of the sky is the same as that parallel to the
line-of-sight.  A hydrodynamic model for the shock was made by
initiating an explosion at the center of a hydrostatic, isothermal
atmosphere with a power-law density profile (see Nulsen et al. 2005
for more details on the shock model). The power-law index for the
density profile of the unshocked gas was determined from the broken
power-law fit to the surface brightness profile.  The surface
brightness profiles and the best-fits of broken power-law density
models in the various sectors are shown in Fig.  \ref{shock_SB.fig},
where the radial axes indicate the distance from the center of
curvature of the annular regions used to extract the profiles.
However, in the following discussion we indicate the position of the
shock in each sector (referred to as the ``shock radius'') in terms of
the distance of the front at mid-aperture of the sector from the
cluster center.  The best-fit model for each sector is summarized in
Table \ref{shock_SB.tab} and the corresponding shock front is shown by
red arcs in Fig. \ref{shock_image.fig}.  We also investigated the
effects of varying the radial binning and center of curvature of the
annular regions used to extract the profile.  We found that the fit
results (i.e., shock radius and Mach number, $\mathcal{M}$) do not
depend strongly on the particular choice of the extraction region nor
on the radial range of the fit, with systematic variations \ltsim 5\%
in each sector.

The shock is clearly visible as a surface brightness jump in all
sectors but the 180-270 sector (i.e. the SE quadrant), with a radius
varying between $\sim$205$''$ in the E-W direction to $\sim$365$''$ in
the N-S direction. The Mach number varies between 1.20 and 1.32 in the
sectors where the density jump is detected, whereas to the SE (sector
180-270) where the surface brightness discontinuity is less evident we
estimate $\mathcal{M} = 1.10$.  These results are consistent with
Nulsen et al. (2005), and partially in agreement with Simionescu et
al. (2009b), who report a detection of the shock front also in the S
direction.


\subsection{Temperature Variations Across the Shock Front}
\label{temp-shock.sec}

Our shock model predicts the emission-weighted temperature to rise
across the front by $\sim$8\%, 10\%, 15\%, 20\% for Mach numbers of
1.18, 1.23, 1.33, 1.40, respectively, reaching its peak at a distance
of $\sim$5-10\% of the shock radius behind the shock and declining
below the unshocked temperature values inside radial distances of
$\sim$25\% of the shock radius.  We used this information to optimize
our selection of the pre-shock and post-shock regions in each sector
and then extracted the spectra in these regions using the {\ttfamily
  SPECEXTRACT} task, which also computes the corresponding
event-weighted response matrices.  Spectral fitting to a single
absorbed {\ttfamily apec} model was performed in XSPEC version 12.3.1
in the 0.5-8.0 keV energy range. Abundances were measured relative to
the abundances of Anders \& Grevesse (1989) and a galactic hydrogen
column of $4.68 \times 10^{20} {\rm cm}^{-2}$ (Dickey \& Lockman 1990)
was assumed.  As a general result, the post-shock regions are found to
be hotter than the corresponding pre-shock regions, although due to
the large error bars the temperature of the pre-shock and post-shock
gas is consistent. Some examples of these temperature measurements are
presented in Sect. 6.1.  A further attempt to bin sectors together in
order to increase the statistics did not lead to any clear detections
of a temperature jump across the shock front (see Sect. 6.1).
However, previous measurements (David et al. 2001; Simionescu et
al. 2009b) indicate that the underlying global temperature profile has
a peak close to the inner edge of the shock front, thus complicating
the detection of a temperature rise due to the shock itself.  To
investigate this in more detail, we study the properties of the
azimuthally-averaged cluster temperature profile in
Sect. \ref{t_profile.sec}.


\section{Global Cluster Temperature Profile} 
\label{t_profile.sec}


\begin{figure}
\centerline{
\includegraphics[width=9cm]{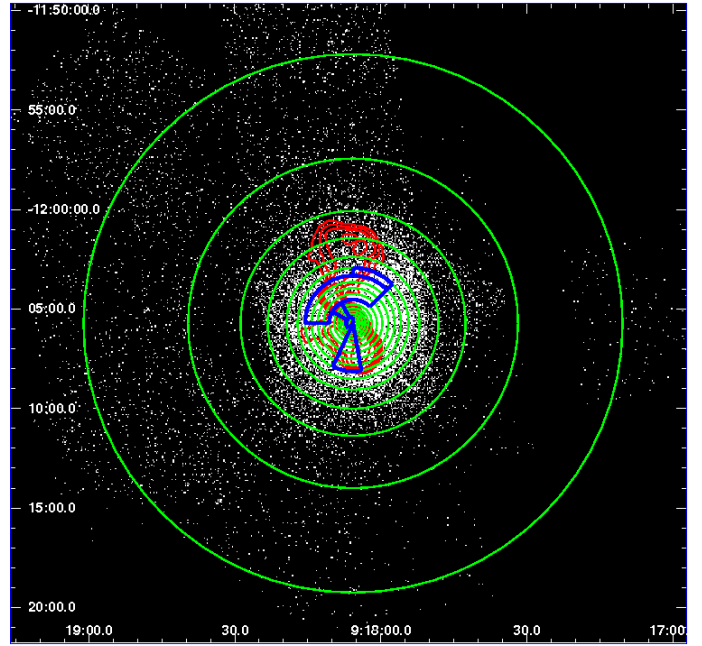}
}
\caption{\small \label{annuli_temp.fig} {\it Chandra} 0.5-7.0 keV mosaic of
  the Hydra A field from Wise et al. (2007). Overlaid in green are the
  annuli used to measure the azimuthally-averaged temperature profile
  shown in the left panel of Fig. \ref{temp.fig}.  The blue sectors
  indicate the regions where the cool filament is present.  These
  sectors have been excluded in determining the temperature profile
  shown in the right panel of Fig. \ref{temp.fig} (see Sect.
  \ref{spectral_cold.sec} for details.).  The red contours outlining
  the 330 MHz radio emission from Lane et al. (2004) are also shown
  for comparison. }
\end{figure}


\begin{figure*}
\centerline{
\includegraphics[width=9cm]{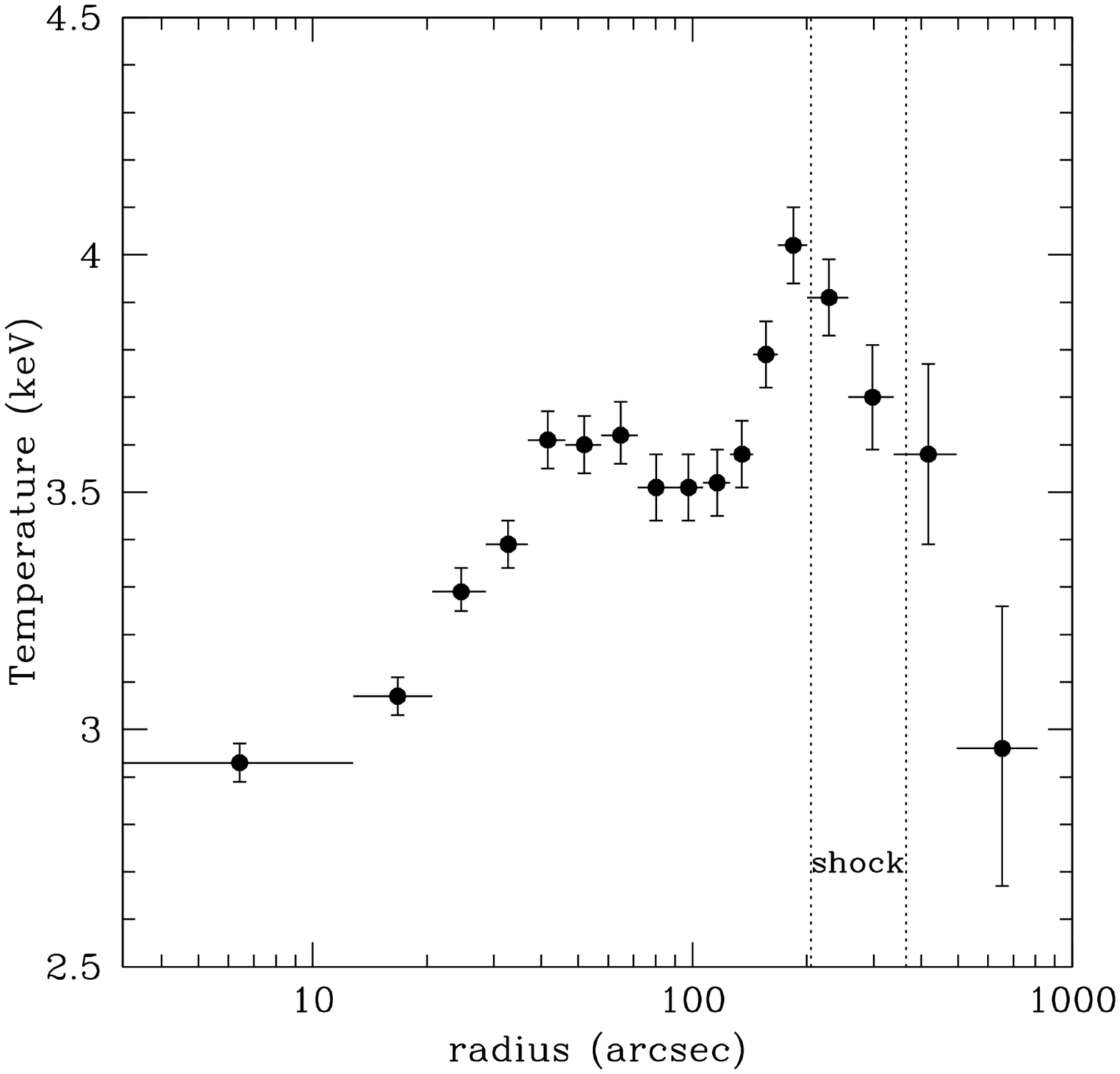}
\includegraphics[width=9cm]{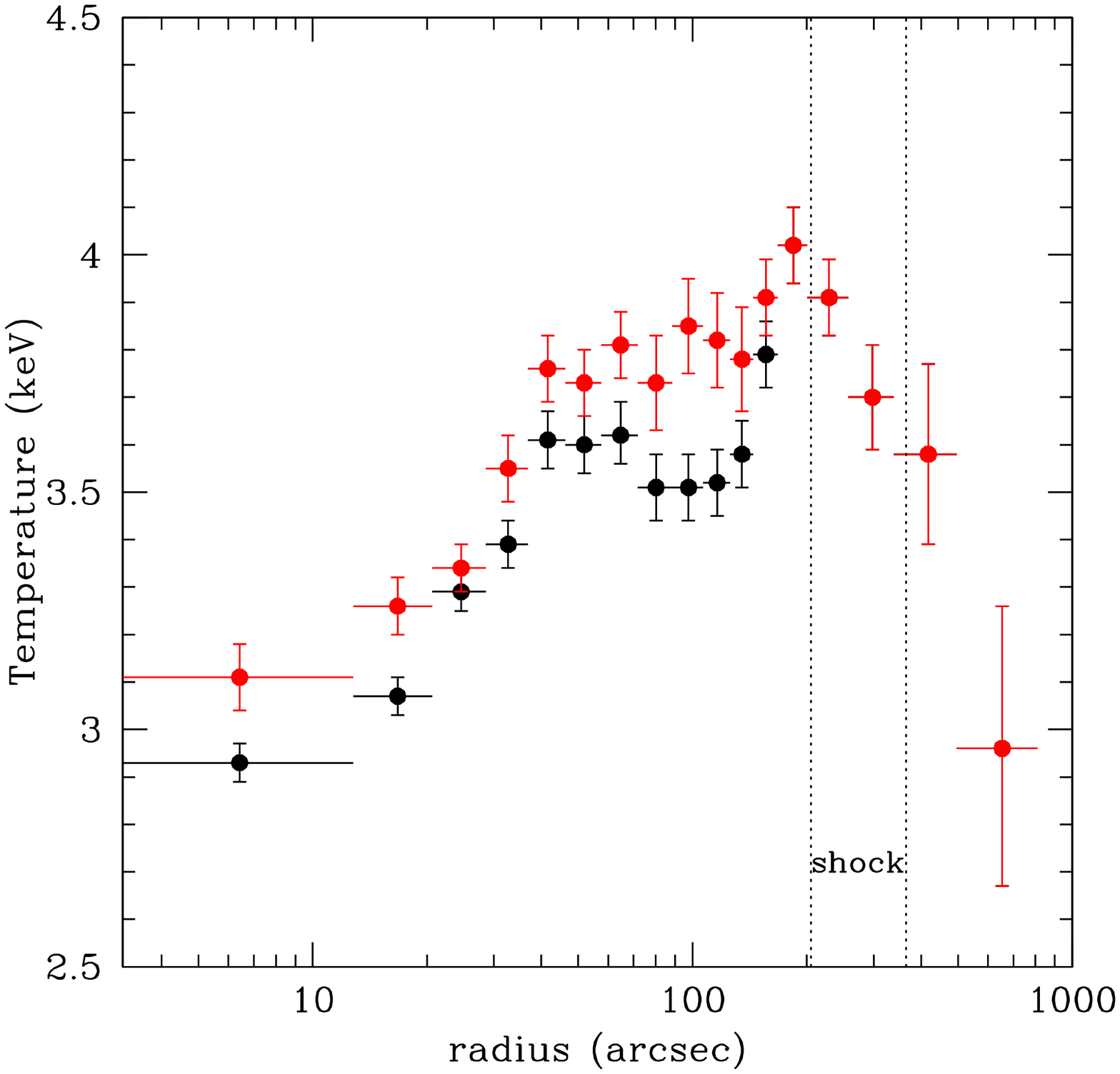}
}
\caption{\small \label{temp.fig} {\it Left:} Azimuthally averaged gas
  temperature profile derived from the ACIS-S and ACIS-I data by
  fitting spectra in the annuli shown in
  Fig. \ref{annuli_temp.fig}. The dotted lines indicate the inner and
  outer radii of the shock front as determined in Sect.
  \ref{azimuthal.sec}.  {\it Right:} Same as left panel, with overlaid
  in red the profile measured after excluding the cool filaments
  (i.e., after excluding the blue sectors shown in
  Fig. \ref{cold.fig}, left panel).}
\end{figure*}



\begin{deluxetable}{cccccc}
\tabletypesize{\scriptsize}
\tablewidth{0pt}
\tablecaption{Temperature Profile derived by Spectral fitting
\label{profile.tab}
} \tablehead{ 
\colhead{Bin} &
\colhead{Radius} & \colhead{$kT$} & \colhead{$\chi^2$/dof}
& \colhead{$kT_{\rm mask}$} &  \colhead{$\chi^2$/dof}\\
\colhead{no.} & \colhead{($''$)} & \colhead{(keV)} & & \colhead{(keV)} } 
\startdata
1 & 0-13 & $2.93^{+0.04}_{-0.04}$ & 763/560 
& $3.11^{+0.07}_{-0.07}$ & 453/430
\\[+1mm]
2 & 13-21 & $3.07^{+0.04}_{-0.04}$ & 638/590 
& $3.26^{+0.06}_{-0.06}$ & 520/466
\\[+1mm]
3 & 21-29 & $3.29^{+0.05}_{-0.04}$ & 607/598 
& $3.34^{+0.05}_{-0.05}$ & 589/566
\\[+1mm]
4 & 29-37 & $3.39^{+0.05}_{-0.05}$ & 663/600 
& $3.55^{+0.07}_{-0.07}$ & 628/565
\\[+1mm]
5 & 37-46 & $3.61^{+0.06}_{-0.06}$ & 681/622 
& $3.76^{+0.07}_{-0.07}$ & 605/584
\\[+1mm]
6 & 46-58 & $3.60^{+0.06}_{-0.06}$ & 644/607 
& $3.73^{+0.07}_{-0.07}$ & 615/578
\\[+1mm]
7 & 58-72 & $3.62^{+0.07}_{-0.06}$ & 737/623 
& $3.81^{+0.07}_{-0.07}$ & 618/593
\\[+1mm]
8 & 72-89 & $3.51^{+0.07}_{-0.07}$ & 656/627 
& $3.73^{+0.10}_{-0.10}$ & 512/502
\\[+1mm]
9 & 89-107 & $3.51^{+0.07}_{-0.07}$ & 748/629 
& $3.85^{+0.10}_{-0.10}$ & 538/513
\\[+1mm]
10 & 107-126 & $3.52^{+0.07}_{-0.07}$ & 698/634 
& $3.82^{+0.10}_{-0.10}$ & 507/508
\\[+1mm]
11 & 126-145 & $3.58^{+0.07}_{-0.07}$ & 713/648
& $3.78^{+0.11}_{-0.11}$ & 491/511
\\[+1mm]
12 & 145-168 & $3.79^{+0.07}_{-0.07}$ & 744/674
& $3.91^{+0.08}_{-0.08}$ & 672/643
\\[+1mm]
13 & 168-201 & $4.02^{+0.08}_{-0.08}$ & 719/709
& $4.02^{+0.08}_{-0.08}$ & 719/709
\\[+1mm]
14 & 201-257 & $3.91^{+0.08}_{-0.08}$ & 1125/1124
& $3.91^{+0.08}_{-0.08}$ & 1125/1124
\\[+1mm]
15 & 257-339 & $3.70^{+0.11}_{-0.11}$ & 1252/1177 
& $3.70^{+0.11}_{-0.11}$ & 1252/1177
\\[+1mm]
16 & 339-497 & $3.58^{+0.19}_{-0.19}$ & 1140/1150 
& $3.58^{+0.19}_{-0.19}$ & 1140/1150
\\[+1mm]
17 & 497-812 & $2.96^{+0.30}_{-0.29}$ & 1717/1676 &
$2.96^{+0.30}_{-0.29}$ & 1717/1676
\\[-2mm]
\enddata
\tablecomments{ Results of the spectral fitting in concentric
  $360^{\circ}$ annular regions (shown in Fig. \ref{annuli_temp.fig})
  in the 0.5-8.0 keV energy range using the XSPEC {\ttfamily
    wabs$\times$apec} model.  The absorbing column density is fixed to
  the Galactic value ($N_{\rm H} = 4.68 \times 10^{20} {\rm cm}^{-2}
  $), while the temperature (in keV) and metallicity (non reported
  here) are left as free parameters. Error bars are at the 90\%
  confidence levels on a single parameter of interest. The values of
  $kT_{\rm mask}$ are measured in the same concentric annular regions
  after excluding the cool filaments (i.e., by fitting spectra
  extracted after masking the blue sectors shown in
  Fig. \ref{cold.fig}, left panel). See Sect. \ref{t_profile.sec} for
  details.}
\end{deluxetable}


The azimuthally averaged gas temperature profile was derived from the
ACIS-S and ACIS-I data by extracting spectra in the annular regions
indicated in Fig. \ref{annuli_temp.fig}, and is shown in the left
panel of Fig. \ref{temp.fig}.  The annular bins and temperature
measurements are detailed in Table \ref{profile.tab}.  The outer two
temperature points represent the farthest temperature measurements
from the cluster center obtained with {\it Chandra} data at present.
We note that the temperature profile peaks around 180$''$, just inside
the inner edge of the shock front. However, as discussed in
Sect. \ref{discussion.sec}, it is unlikely that this temperature
feature is produced by the shock as it is consistent with the general
shape of temperature profiles observed for relaxed galaxy clusters.

On the other hand, a {\it ``plateau''} is notable in the temperature
profile indicating the presence of cool gas in the range of radius
$\sim$70-150$''$. This peculiar feature has not been noted previously
in the literature. We thus investigated it in more detail thanks to
the deep, high-quality {\it Chandra} exposure.


\subsection{Evidence for Cool Filaments}

In order to investigate the origin of the plateau seen in the
temperature profile between $\sim$ 70-150$''$, we made several
attempts to identify regions of cool gas.  Our approach was to measure
a new, ``undisturbed'' temperature profile by masking such regions and
then compare it with the global temperature profile measured above.
We first masked the cavity regions (as indicated in Wise et al. 2007),
finding systematically higher temperatures by $\sim$0.1-0.2 keV in all
bins (except for the very central bin, where the temperature is lower
by $\sim$0.15 keV) up to $\sim$200$''$ . The temperature profile
measured by masking the cavities is therefore shifted up, maintaining
its global shape. In particular, the plateau in the temperature
profile is still present, indicating that the cavities are not a
tracer of the entire amount of cool gas.  We then masked the
$\sim$1$'$ bright filament stretching from the inner cavity to the
center of a larger outer cavity in the northeast, which was noted by
Nulsen et al. (2005). As in the previous attempt, the plateau in the
temperature profile remained, indicating that the known filament is
not the only repository of the cool gas in Hydra A.

As a diagnostic of the presence of cool gas we finally used the
hardness ratio map shown in Fig. \ref{cold.fig}, which was obtained by
dividing a smoothed 1.5-7.5 keV image by a smoothed 0.3-1.5 keV image.
The dark regions are indicative of low gas temperatures. They agree
with the arm-like structures of cooler gas extending towards the north
and the south in the temperature map of Simionescu et al. (2009a).
Based on a visual inspection of the hardness ratio map, we selected a
simple combination of sectors reproducing the shape of the supposedly
cool filaments (see left panel of Fig. \ref{cold.fig}).  We then
excluded these regions and extracted the spectra in the same radial
bins as above (see Fig.  \ref{annuli_temp.fig}), and generated the
temperature profile shown in red in the right panel of
Fig. \ref{temp.fig} (see Table \ref{profile.tab}).  As evident from
the comparison of the two profiles, the plateau has largely been
removed and the temperature profile is typical of cool core clusters
(see also Sect.  \ref{discussion.sec}).  This clearly indicates that
the masked regions contain cool gas.  A detailed spectral analysis of
the cool filaments is presented in Sect. \ref{spectral_cold.sec}.


\section{Spectral Properties of the Cool Gas}
\label{spectral_cold.sec}


\begin{figure*}
\centerline{
\includegraphics[width=9cm]{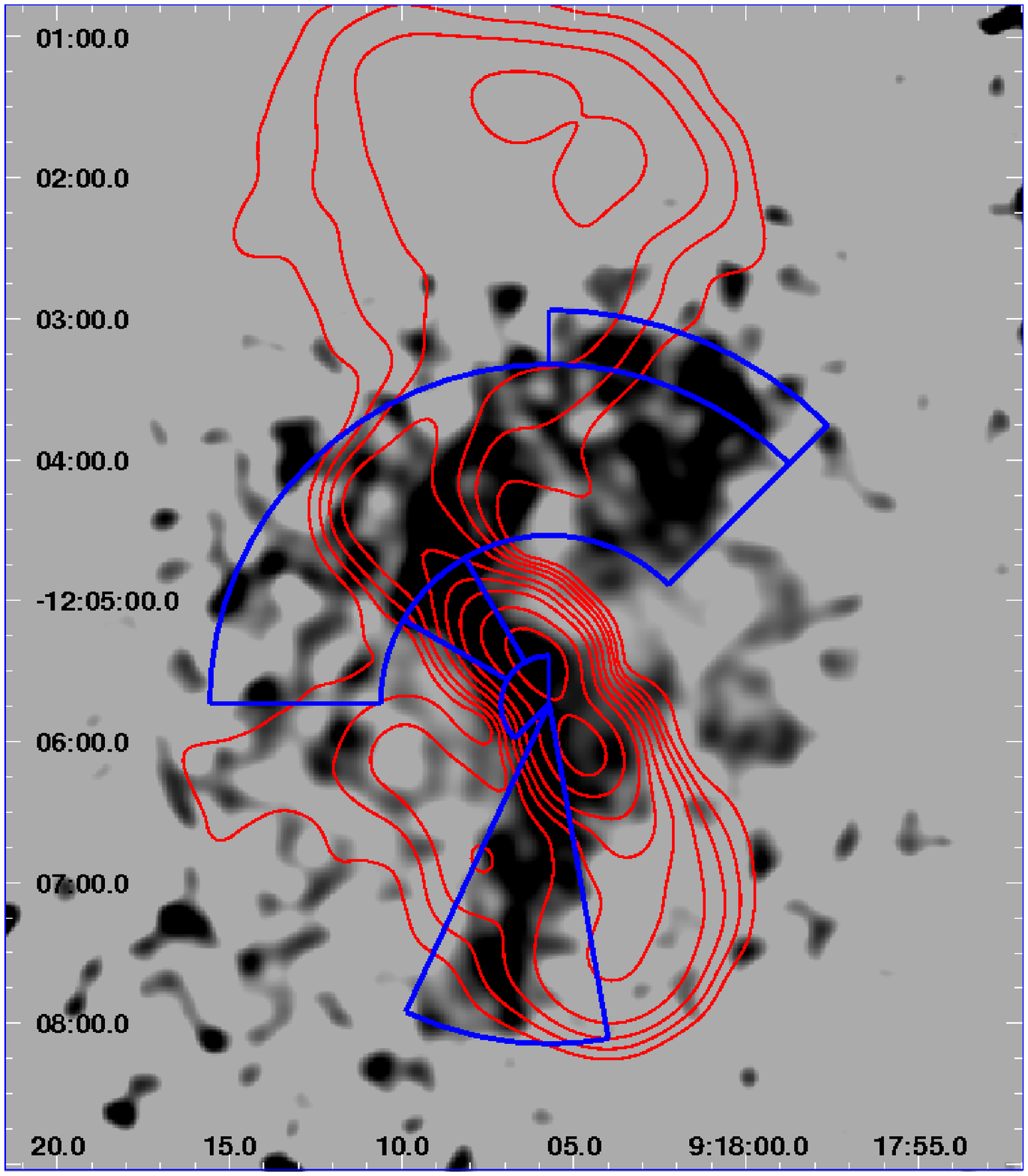}
\includegraphics[width=9cm]{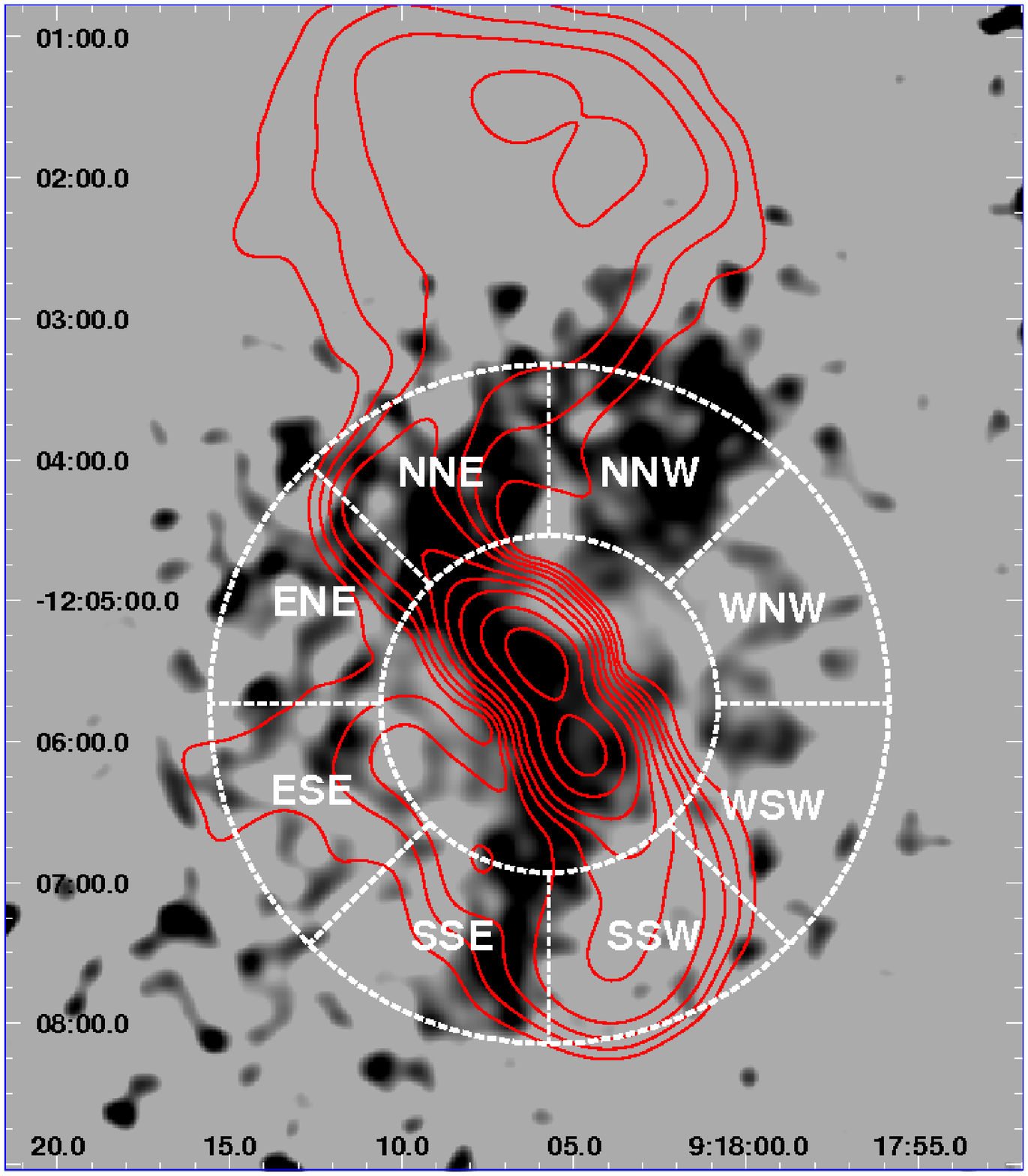}
}
\caption{\small \label{cold.fig} {\it Left:} High-contrast hardness ratio map
  obtained by dividing a 1.5-7.5 keV image by a 0.3-1.5 keV image.
  Each raw image was smoothed with a 10 pixel (5$''$) gaussian before
  the division.  The images were extracted from the merged cleaned
  event files for ObsID's 4969 and 4970. Regions in black are
  indicative of low temperature gas.  The blue sectors indicate the
  selected regions of supposedly cool filaments. These regions have
  been excluded in measuring the temperature profile shown in the
  right panel of Fig. \ref{temp.fig} (red points).  The red contours
  outlining the 330 MHz radio emission from Lane et al. (2004) are
  also shown for comparison. {\it Right:} Similar to left
  panel. Overlaid in white are the sectors used to study the spectral
  properties of the cool filaments (located between radii 72-145$''
  \sim$ 76-152 kpc, see Sect.  \ref{spectral_cold.sec} for details).
}
\end{figure*}


\begin{deluxetable*}{cccccccccccc}
\tablewidth{0pt}
\tabletypesize{\scriptsize}
\tablecaption{Spectral Analysis of the Cool Filaments
\label{spectral_study.tab}
}
\tablehead{
\colhead{} & \multicolumn{4}{c}{1T model} &  \multicolumn{7}{c}{2T model}\\
\colhead{Sector} & \colhead{$kT$ (keV)} & \colhead{EM $(\times 10^{-3})$}  & \colhead{$Z$ (solar)} & \colhead{$\chi^2$/dof [$\chi^2_{\nu}$]} & \colhead{$kT_1$ (keV)}  & \colhead{EM$_1 (\times 10^{-3})$}  & \colhead{$kT_2$(keV)} & \colhead{EM$_2 (\times 10^{-3})$ } & \colhead{$Z$ (solar)} & \colhead{$\chi^2$/dof} & \colhead{F stat} }
\startdata
WNW & $3.87^{+0.10}_{-0.10}$ & $1.60^{+0.03}_{-0.03}$  & $0.37^{+0.05}_{-0.05}$
& 505.1/521 [0.97]
& $2.17^{+0.48}_{-0.47}$  & $0.73^{+0.28}_{-0.45}$
& $6.40^{+1.89}_{-1.74}$  & $0.94^{+0.43}_{-0.28}$  & $0.28^{+0.06}_{-0.05}$
& 484.6/519 & 11.0
\\[+1mm]
NNW  & $3.24^{+0.06}_{-0.07}$ & $1.79^{+0.03}_{-0.03}$  & $0.36^{+0.04}_{-0.04}$
& 524.9/525 [1.00] 
& $2.05^{+0.17}_{-0.25}$  & $1.08^{+0.24}_{-0.34}$  
& $6.74^{+4.01}_{-1.75}$  & $0.80^{+0.31}_{-0.35}$ & $0.26^{+0.04}_{-0.04}$
& 467.6/523 & 32.0
\\[+1mm]
NNE  & $3.14^{+0.08}_{-0.08}$ & $1.23^{+0.03}_{-0.03}$  & $0.41^{+0.05}_{-0.05}$
& 483.0/460 [1.05]
& $2.12^{+0.23}_{-0.21}$  & $0.81^{+0.06}_{-0.29}$  
& $6.84^{+1.48}_{-2.20}$  & $0.48^{+0.26}_{-0.11}$ & $0.30^{+0.05}_{-0.05}$
& 451.9/458 & 15.8
\\[+1mm]
ENE  & $3.32^{+0.08}_{-0.08}$ & $1.15^{+0.02}_{-0.02}$  & $0.42^{+0.06}_{-0.05}$
& 455.5/455 [1.00] 
& $2.24^{+0.37}_{-0.47}$  & $0.72^{+0.20}_{-0.47}$ 
& $6.84^{+4.62}_{-2.72}$  & $0.48^{+0.50}_{-0.20}$ & $0.34^{+0.06}_{-0.05}$
& 433.6/453 & 11.4
\\[+1mm]
ESE  & $3.45^{+0.10}_{-0.10}$ & $1.36^{+0.03}_{-0.03}$  & $0.37^{+0.05}_{-0.05}$
& 502.2/483 [1.04]
& $2.16^{+0.86}_{-0.48}$  & $0.48^{+0.89}_{-0.35}$
& $4.49^{+14.9}_{-0.61}$  & $0.92^{+0.33}_{-0.66}$  & $0.32^{+0.05}_{-0.05}$
& 491.2/481 & 5.4
\\[+1mm]
SSE  & $3.34^{+0.07}_{-0.07}$ & $1.45^{+0.03}_{-0.03}$  & $0.38^{+0.05}_{-0.05}$
& 621.8/491 [1.27] 
& $1.59^{+0.11}_{-0.23}$  & $0.44^{+0.18}_{-0.21}$
& $4.92^{+0.97}_{-0.74}$  & $1.05^{+0.17}_{-0.15}$  & $0.28^{+0.06}_{-0.05}$
& 533.9/489 & 40.3
\\[+1mm]
SSW & $4.01^{+0.12}_{-0.12}$ & $1.18^{+0.02}_{-0.02}$  & $0.35^{+0.06}_{-0.06}$
& 458.0/472 [0.97]
& $2.39^{+0.65}_{-0.90}$  & $0.67^{+0.17}_{-0.53}$  
& $8.60^{+4.98}_{-4.12}$  & $0.58^{+0.50}_{-0.22}$ & $0.28^{+0.07}_{-0.07}$
& 438.2/470 & 10.6
\\[+1mm]
WSW & $3.92^{+0.11}_{-0.11}$ & $1.37^{+0.03}_{-0.03}$  & $0.31^{+0.06}_{-0.05}$
& 439.4/494 [0.89]
& $2.93^{+0.33}_{-1.14}$  & $1.17^{+0.07}_{-0.85}$  
& $54.4^{+/}_{-/}$ & $0.36^{+0.12}_{-0.12}$ & $0.25^{+0.06}_{-0.05}$
& 419.5/492 & 11.7 
\\[-2mm]
\enddata
\tablecomments{ Results of 1T model and 2T model spectral fitting in
  the 0.5-8.0 keV energy range. The annular sectors of spectra
  extraction are indicated in the right panel of
  Fig. \ref{cold.fig}. The normalizations (EMs) are in XSPEC units of
  $10^{-14} n_{\rm e} n_{\rm p} V / 4 \pi [D_{\rm A} (1+z)]^2$. Errors
  are at the 90\% confidence levels on a single parameter of
  interest. The last column shows the result of the F-test. See
  Sect. \ref{spectral_cold.sec} for details.  }
\end{deluxetable*}


\begin{figure*}
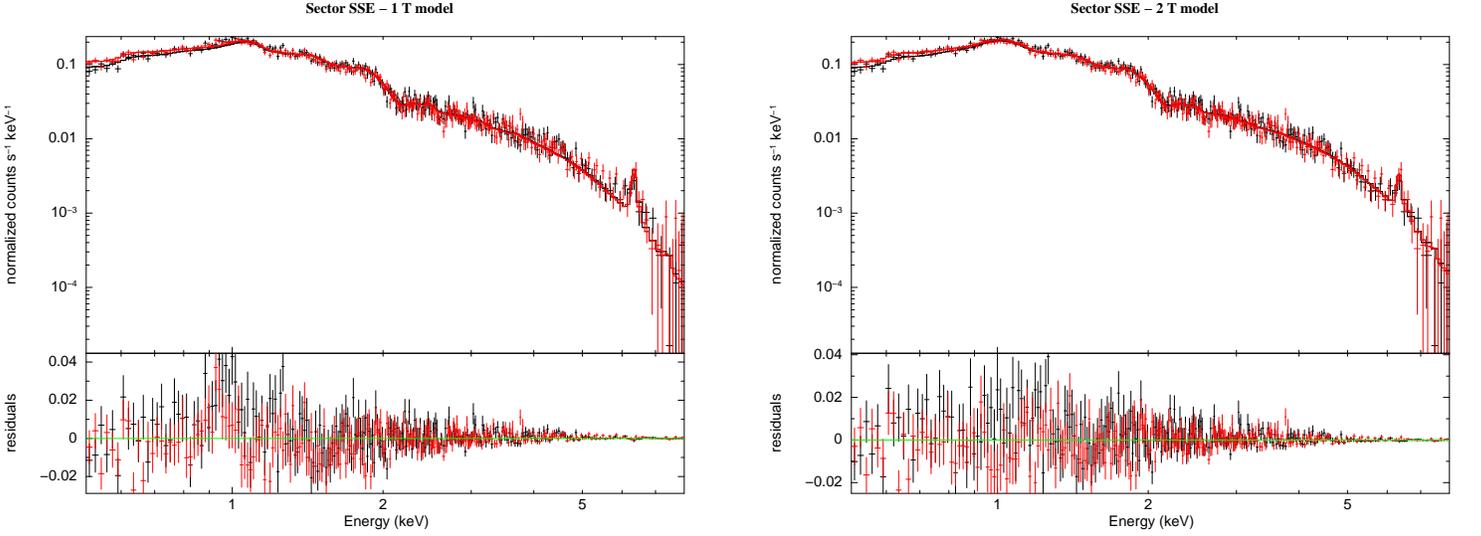

\centerline{
\includegraphics[width=7.5cm, angle=-90]{gitti_fig6a.eps}
\includegraphics[width=7.5cm, angle=-90]{gitti_fig6b.eps}
}
\caption{\small \label{spectra.fig} 1T model (left) and 2T model (right) fit
  to the spectra extracted in the SSE sector. See Table
  \ref{spectral_study.tab} for best-fitting parameter values.}
\end{figure*}


\begin{figure}
\centerline{
\includegraphics[width=9cm]{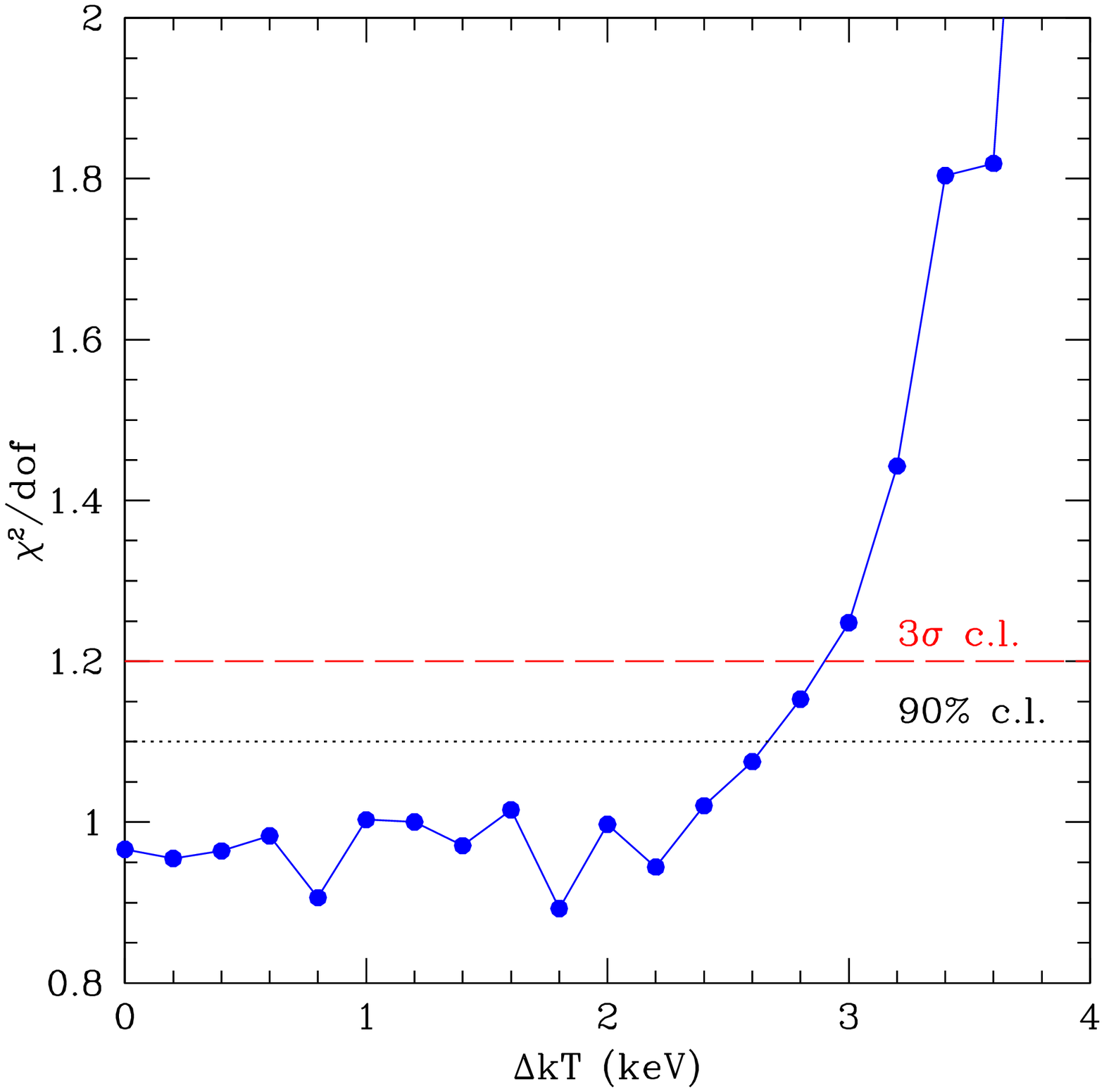}
}
\caption{\small \label{EM_distrib.fig} 
  Reduced chi square of the 1T model fit to mock 2-temperature spectra
  as a function of the separation $\Delta$T of the 2 thermal
  components. The best-fit value of the single temperature fit is 
  $kT = 3.24 \pm 0.06$ keV.  
}
\end{figure}


We investigate here the spectral properties of the gas which produces
the plateau seen in the global cluster temperature profile.  In
particular, we focus on the gas located in the range of projected
distances from the center of 72-145$''$ ($\sim$ 76-152 kpc), which
corresponds to the radial range where the plateau in the temperature
profile is most evident (i.e., bins no. 8-11, see Fig. \ref{temp.fig}
and Table \ref{profile.tab}).  We divided the annulus from 72-145$''$
into 8 sectors, each having an angular width of $45^{\circ}$,
obtaining the regions labeled as WNW (west-northwest), NNW
(north-northwest), NNE (north-northeast), ENE (east-northeast), ESE
(east-southeast), SSE (south-southeast), SSW (south-southwest), WSW
(west-southwest) in the right panel of Fig. \ref{cold.fig}.  We
extracted the spectra in these sectors and compared two different
spectral models. The ``1T model'' is the absorbed {\ttfamily apec}
model already used above to derive the global temperature profile.
The free parameters are the temperature, $kT$, the metallicity, $Z$,
and the normalization (emission measure, EM).  The ``2T model''
includes a second thermal emission component ({\ttfamily apec+apec})
and has 2 additional free parameters: the temperature, $kT_2$, and the
normalization, EM$_2$, of the second component (the metallicities of
the two components are linked).


\begin{deluxetable}{cccccc}
\tabletypesize{\scriptsize}
\tablewidth{0pt}
\tablecaption{Spectral Analysis of the Cool Filaments (cont.)
\label{spectral_study2.tab}
} \tablehead{
  \colhead{} & 
  \multicolumn{5}{c}{2T model with $kT_2 = 4$ keV}\\
  \colhead{Sector} & 
   \colhead{$kT_1$ (keV)} & \colhead{EM$_1
    (\times 10^{-3})$} & \colhead{EM$_2 (\times 10^{-3})$ } &
   \colhead{$\chi^2$/dof} & \colhead{F stat}} 
  \startdata
WNW 
& $1.33^{+0.37}_{-0.24}$  & $0.05^{+0.05}_{-0.04}$ & $1.55^{+0.03}_{-0.04}$ 
& 493.8/520 & 11.9
\\[+1mm]
NNW  
& $1.67^{+0.16}_{-0.11}$  & $0.39^{+0.08}_{-0.07}$ & $1.42^{+0.06}_{-0.07}$ 
& 478.8/524 & 50.5
\\[+1mm]
NNE  
& $1.73^{+0.28}_{-0.08}$  & $0.33^{+0.10}_{-0.05}$ & $0.93^{+0.04}_{-0.10}$ 
& 456.9/459 & 26.2
\\[+1mm]
ENE  
& $1.72^{+0.35}_{-0.13}$  & $0.22^{+0.09}_{-0.05}$ & $0.95^{+0.05}_{-0.09}$ 
& 436.6/454 & 19.7
\\[+1mm]
ESE  
& $2.00^{+0.52}_{-0.35}$  & $0.27^{+0.17}_{-0.09}$ & $1.11^{+0.09}_{-0.17}$ 
& 492.5/482 & 9.5
\\[+1mm]
SSE  
& $1.33^{+0.12}_{-0.08}$  & $0.18^{+0.05}_{-0.04}$ & $1.26^{+0.03}_{-0.03}$ 
& 539.1/490 & 75.2
\\[+1mm]
SSW 
& $1.09^{+0.48}_{-0.26}$  & $0.02^{+0.03}_{-0.01}$ & $1.17^{+0.03}_{-0.02}$ 
& 451.4/471 & 6.9
\\[+1mm]
WSW 
& $1.37^{+0.63}_{-0.35}$  & $0.04^{+0.05}_{-0.03}$ & $1.34^{+0.03}_{-0.04}$ 
& 433.1/493 & 7.2
\\[-2mm]
\enddata
\tablecomments{ Results of 2T model spectral fitting in the 0.5-8.0
  keV energy range. The temperature value of the second thermal
  component is fixed to 4 keV.  The annular sectors of spectra
  extraction are indicated in the right panel of
  Fig. \ref{cold.fig}. The normalizations (EMs) are in XSPEC units of
  $10^{-14} n_{\rm e} n_{\rm p} V / 4 \pi [D_{\rm A} (1+z)]^2$. Errors
  are at the 90\% confidence levels on a single parameter of
  interest. The last column shows the F statistic value testing the
  improvement of the 2T model over the 1T model in Table
  \ref{spectral_study.tab}. See Sect. \ref{spectral_cold.sec} for
  details.  }
\end{deluxetable}


The best-fitting parameter values and 90\% confidence ranges derived
from the fits to the annular spectra in sectors are summarized in
Table \ref{spectral_study.tab}.  Although the improvement of adding a
second thermal component is formally significant according to the
F-test, our results show that the quality of the data is not generally
sufficient to demand a model more complex than the 1T model.
In fact, in most sectors, the 1T model already produces a very good
fit (reduced chi squared $\chi^2_{\nu} \sim 1$) and therefore a more
complicated model appears unnecessary.  We also note that the second
thermal component is poorly constrained, with temperature errors
\gtsim 25\% and up to 300\%.  Only in the SSE sector is the reduced
chi squared of the 1T model unacceptable at 90\% significance, and
the statistical improvement obtained by introducing an additional
emission component compared to the single-temperature model is the
most significant according to the F-test.  The improvement of the 2T
model over the 1T model in this sector is also evident from
the residuals of the fits in Fig. \ref{spectra.fig}.  We can therefore
conclude that, confirming the hardness ratio map (Fig. \ref{cold.fig},
right panel), we find spectral evidence for multiphase gas in the SSE
sector with a hot component at $4.92^{+0.97}_{-0.74}$ keV and a cool
component at $1.59^{+0.11}_{-0.23}$ keV.  Assuming that the two
spectral phases are in pressure equilibrium in the same volume, the
ratio of the volumes they occupy is estimated as $V_1/V_2 = ({\rm
  EM}_1/{\rm EM}_2) \cdot (kT_1/kT_2)^2 $, so the filling factor of
the cool gas is $\sim$0.04.


\begin{figure*}
\centerline{
\includegraphics[width=9cm]{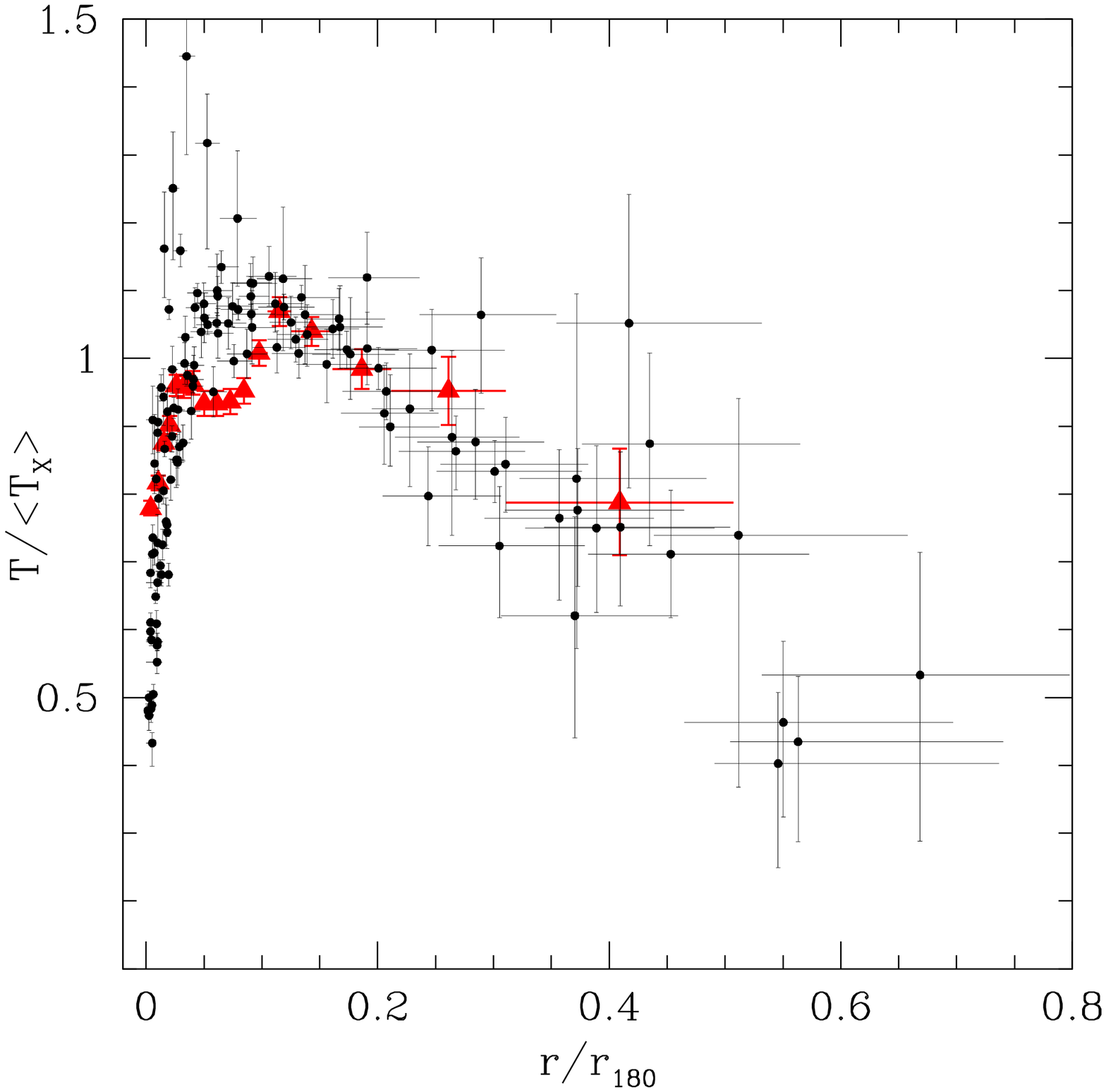}
\includegraphics[width=9cm]{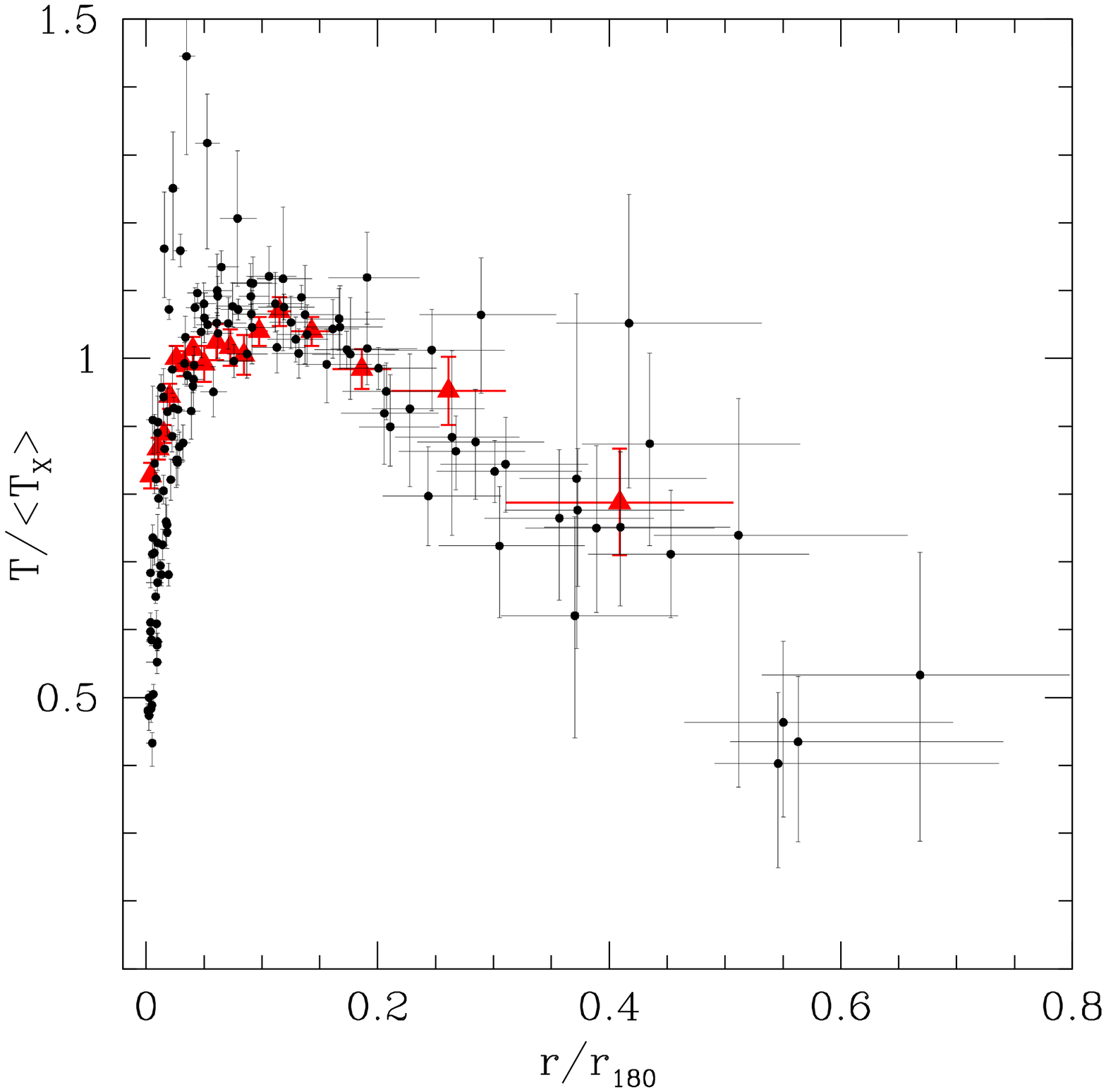}
}
\caption{\small \label{scaled_temp.fig} {\it Left:} Temperature profile
  measured for Hydra A (red triangles, corresponding to the profile
  shown in the left panel of Fig. \ref{temp.fig}) overlaid on the
  temperature profiles of a sample of 12 relaxed clusters presented by
  Vikhlinin et al. (2005).  The temperatures are scaled to the cluster
  emission-weighted temperature excluding the central 70 kpc regions.
  By extracting the global spectrum of Hydra A, after masking the
  central 67$''$, we measured a value $<T_{\rm X}> = 3.76 \pm 0.03$
  keV. The profiles for all clusters are projected and scaled in
  radial units of the virial radius $r_{vir}$, estimated from the
  relation $r_{\rm vir} = 2.74 \, {\rm Mpc} \sqrt{<T_{\rm X}>/10 \,
    {\rm keV}}$ (Evrard et al. 1996).  {\it Right:} Similar to the
  left panel, but with the temperature profile measured for Hydra A
  after masking the cool filaments (corresponding to the profile shown
  in the right panel of Fig. \ref{temp.fig}).}
\end{figure*}


By contrast, we do not find clear spectral signatures of cool gas in
the sectors NNW, NNE, and ENE, as expected from a visual inspection of
the hardness ratio map.  However, the lack of spectral evidence for
multiphase gas could be due to the limitations of our data.  Indeed,
due to the relatively limited spectral resolution of Chandra, the
detection of two different thermal components demands a significant
temperature separation.  The temperature difference required to have a
marked effect on a single-phase thermal fit at the 90\% confidence
level is determined as follows.  Using the response matrices,
background, and numerical information of a real spectrum extracted in
an arbitrary sector, we simulated spectra with two thermal components
separated by $\Delta$T around the best-fit value of the single
temperature fit to the real spectrum.  We then fitted a single
temperature {\ttfamily apec} model to the mock 2-temperature
spectrum. We repeated this exercise, increasing the separation between
the two temperature components until $\chi^2_{\nu}$ exceeds the 90\%
confidence range.  For comparison, we performed this procedure by
starting from the real spectra in different sectors indicated in
Fig. \ref{cold.fig} and found consistent results.  An example of such
an analysis is plotted in the left panel of Fig. \ref{EM_distrib.fig},
where the dotted and dashed lines are the 90\% and $3 \sigma$ limits.
We found that a $\sim$2.8 keV (3.0 keV) separation is necessary to
exclude the presence of single-phase gas at the 90\% ($3 \sigma$)
confidence for our data.  The fact that the two spectral components
detected in sector SSE are separated by $\sim$3.3 keV is consistent
with this result.  We note that $kT_1$ found by the 2T model fit in
the SSE sector is the lowest among all of the sectors, falling in the
energy range which comprises most of the counts and thus being more
easily detectable.

Despite this, since a temperature higher than 4 keV is not observed at
any radius in the cluster (Fig. \ref{temp.fig}), the second thermal
component found by the 2T model fit appears unrealistically hot so we
performed a new spectral fit with the 2T model ({\ttfamily apec+apec})
keeping the temperature of the second thermal component fixed at 4
keV.  Such a model has only one additional free parameter than the 1T
model: the normalization, EM$_2$, of the second component (the
metallicities of the two components are linked).  The best-fitting
parameter values and 90\% confidence levels derived from the fits to
the annular spectra in sectors are summarized in Table
\ref{spectral_study2.tab}.  The F statistics for the improvement over
the 1T model, shown in the last column, indicate where the addition of
a second thermal component is most significant (sectors SSE, NNW, NNE,
and ENE).  In agreement with the hardness ratio map
(Fig. \ref{cold.fig}), our spectral analysis therefore supports the
presence of multiphase gas along the filaments.  Interestingly, such
cool filaments follow the morphology of the powerful central radio
source nicely, although the western part of the southern radio lobe
appears devoid of cool gas. The SSW sector is indeed the hottest and
is also the region where the presence of a second thermal component is
least significant. This sector lies at the location where the southern
radio lobe appears to fold back on itself (Lane et al. 2004). The
properties of the cavity are also consistent with a sharp bend in the
southern jet there (Wise et al. 2007).  

Finally, we also attempted to map the emission measure distribution by
fitting more complicated multi-phase spectral models (such as
{\ttfamily apec+apec+apec+apec}, with fixed temperatures and free
normalizations). However, we found our data are inadequate due to
limited statistics and the spectral resolution of {\it Chandra}.
These detailed spectral studies will, hopefully, be possible in the
future with the spectral capabilities of the {\it International X-ray
  Observatory (IXO)}.


\section{Discussion}
\label{discussion.sec}

\subsection{Scaled Temperature Profile and Shock Front}


\begin{deluxetable*}{lccccccccccc}
\tabletypesize{\scriptsize}
\tablewidth{0pt}
\tablecaption{Deprojection analysis
\label{deproj.tab}
} \tablehead{
  \colhead{Shell} 
& \colhead{$kT_{\rm cool}$} & \colhead{EM$_{\rm cool}$} 
& \colhead{$kT_{\rm hot}$} & \colhead{EM$_{\rm hot}$} 
& \colhead{$V_{\rm tot}$} & \colhead{$f_{\rm cool}$} 
& \colhead{$n_{\rm e,cool}$} & \colhead{$M_{\rm cool}$} 
& \colhead{$f_{\rm hot}$} & \colhead{$n_{\rm e,hot}$} & \colhead{$M_{\rm hot}$} 
\\ 
\colhead{($''$)} 
& \colhead{(keV)} & \colhead{$(\times 10^{-3})$} 
& \colhead{(keV)} & \colhead{$(\times 10^{-3})$} & \colhead{(cm$^3$)} 
& \colhead{} & \colhead{$(\rm cm^{-3})$} & \colhead{$({\rm M}_{\odot})$} 
& \colhead{} & \colhead{$(\rm cm^{-3})$} & \colhead{$({\rm M}_{\odot})$} 
}
\startdata 
Inner 
& $1.77^{+0.13}_{-0.06}$ & $3.07$ 
& $4.01^{+0.04}_{-0.04}$ & $13.3$ & $5.35 \times 10^{70}$
& $0.043$ & $0.032$ & $6.86 \times 10^{10}$ 
& $0.957$ & $0.014$ & $6.68 \times 10^{11}$\\ 
(0-72) 
\\
Middle
& $1.58^{+0.08}_{-0.10}$ & $1.64$ 
& $4.01^{+0.04}_{-0.04}$ & $10.9$ & $3.85 \times 10^{71}$
& $0.023$ & $0.012$ & $9.90 \times 10^{10}$ 
& $0.977$ & $0.0047$ & $1.65 \times 10^{12}$\\ 
(72-145)
\\
Outer 
& $-$ & $0$ 
& $4.01^{+0.04}_{-0.04}$ & $12.5$ & $5.20 \times 10^{72}$ 
& $0$ & $0$ & $0$ 
& $1$ & $0.0136$ & $6.59 \times 10^{12}$\\ 
(145-339)
\\[-2mm]
\enddata
\tablecomments{ Results of the deprojection analysis in three
  concentric $360^{\circ}$ annular regions (indicated in the first
  column) in the 0.5-8.0 keV energy range using the XSPEC {\ttfamily
    projct$\times$wabs$\times$(apec+apec)} model. The normalization of
  the cool component in the outer shell is fixed to zero, and the
  temperature values of the hot components are linked between all
  shells.  The fit gives $\chi^2$/dof = 4993.5/2811.  The
  normalizations, EM$_{\rm cool(hot)}$, are in XSPEC units of
  $10^{-14} n_{\rm e,cool(hot)} n_{\rm p} V_{\rm cool(hot)} / 4 \pi
  [D_{\rm A} (1+z)]^2$. Errors are at the 90\% confidence levels on a
  single parameter of interest. The ratio of the relative volumes
  occupied is estimated as $V_{\rm cool}/V_{\rm hot} = ({\rm EM}_{\rm
    cool}/{\rm EM}_{\rm hot}) \cdot (kT_{\rm cool}/kT_{\rm hot})^2 $,
  and the filling factor $f$ is defined as $f_{\rm cool(hot)} = V_{\rm
    cool(hot)}/V_{\rm tot}$, with $V_{\rm tot} = V_{\rm cool} + V_{\rm
    hot}$. The masses are thus estimated from the relation: $M = \rho
  f V_{\rm tot}$, where we use the conversion from electron number
  density to gas density $\rho = 1.83 \mu n_e m_H$.}
\end{deluxetable*}
 

We show in the left panel of Fig. \ref{scaled_temp.fig} the observed
temperature profile of Hydra A (red triangles), scaled by the virial
radius (estimated from the relation $r_{\rm vir} = 2.74 \, {\rm Mpc}
\sqrt{<T_{\rm X}>/10 \, {\rm keV}}$, Evrard et al. 1996) , overlaid on
the scaled temperature profiles of a sample of 12 relaxed clusters
observed with \textit{Chandra} (Vikhlinin et al. 2005).  The
temperatures are scaled to the emission-weighted cluster temperature
$<T_{\rm X}>$, measured after excluding the central 70 kpc region
which is usually affected by radiative cooling.  We note that the
plateau of cool gas, already discussed in Sect.  \ref{t_profile.sec},
stands out from to the typical temperature profile of clusters.  In
particular, lying just inside the typical temperature peak, such a
plateau emphasizes the temperature peak in Hydra A, making it look
higher than it actually is.  Since this region is also where the shock
front is located, it makes it difficult to distinguish between a
temperature jump due to the shock and the typical temperature peak of
the cluster.


\begin{figure*}
\centerline{
\includegraphics[width=7.5cm]{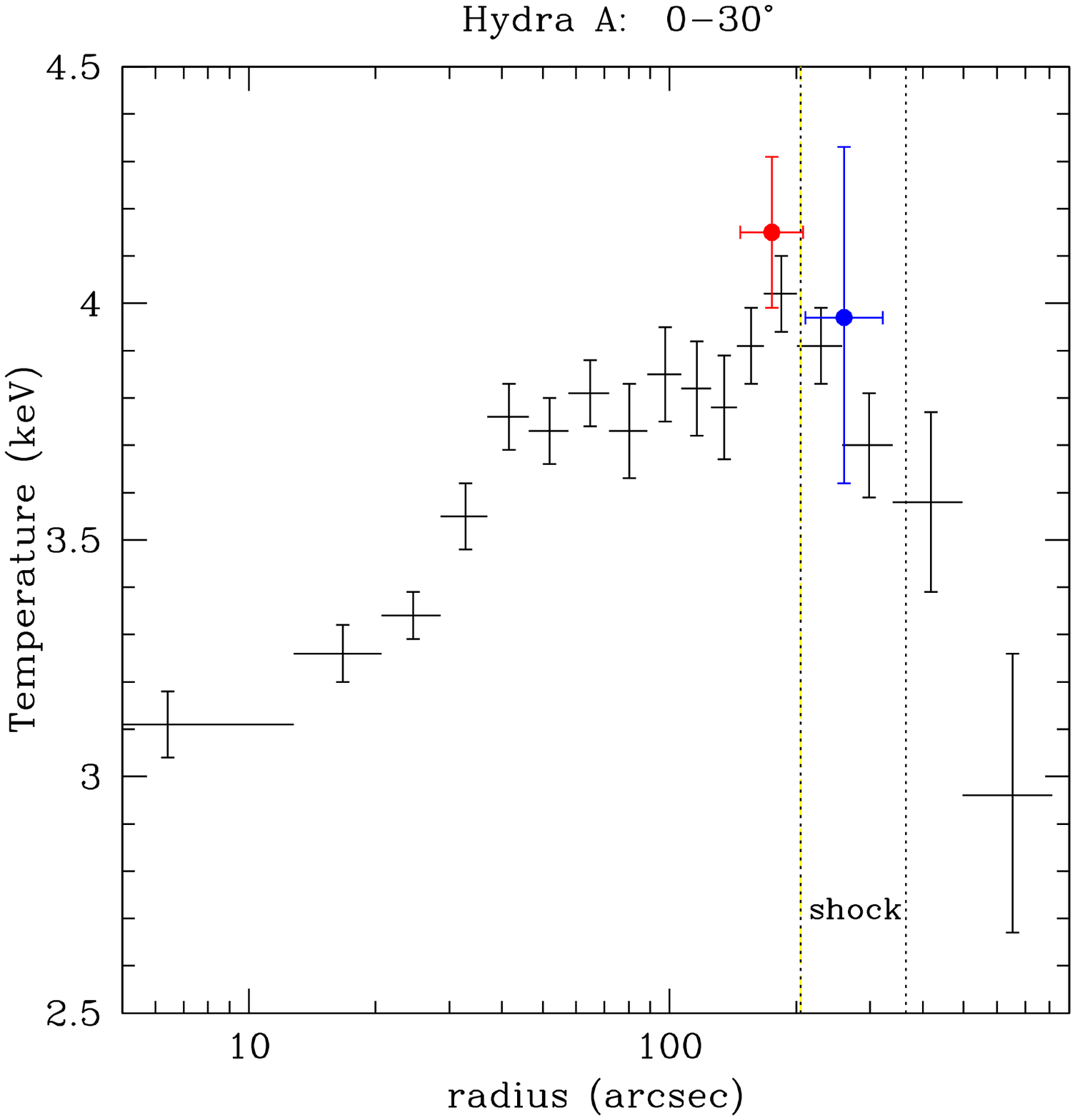}
\includegraphics[width=7.5cm]{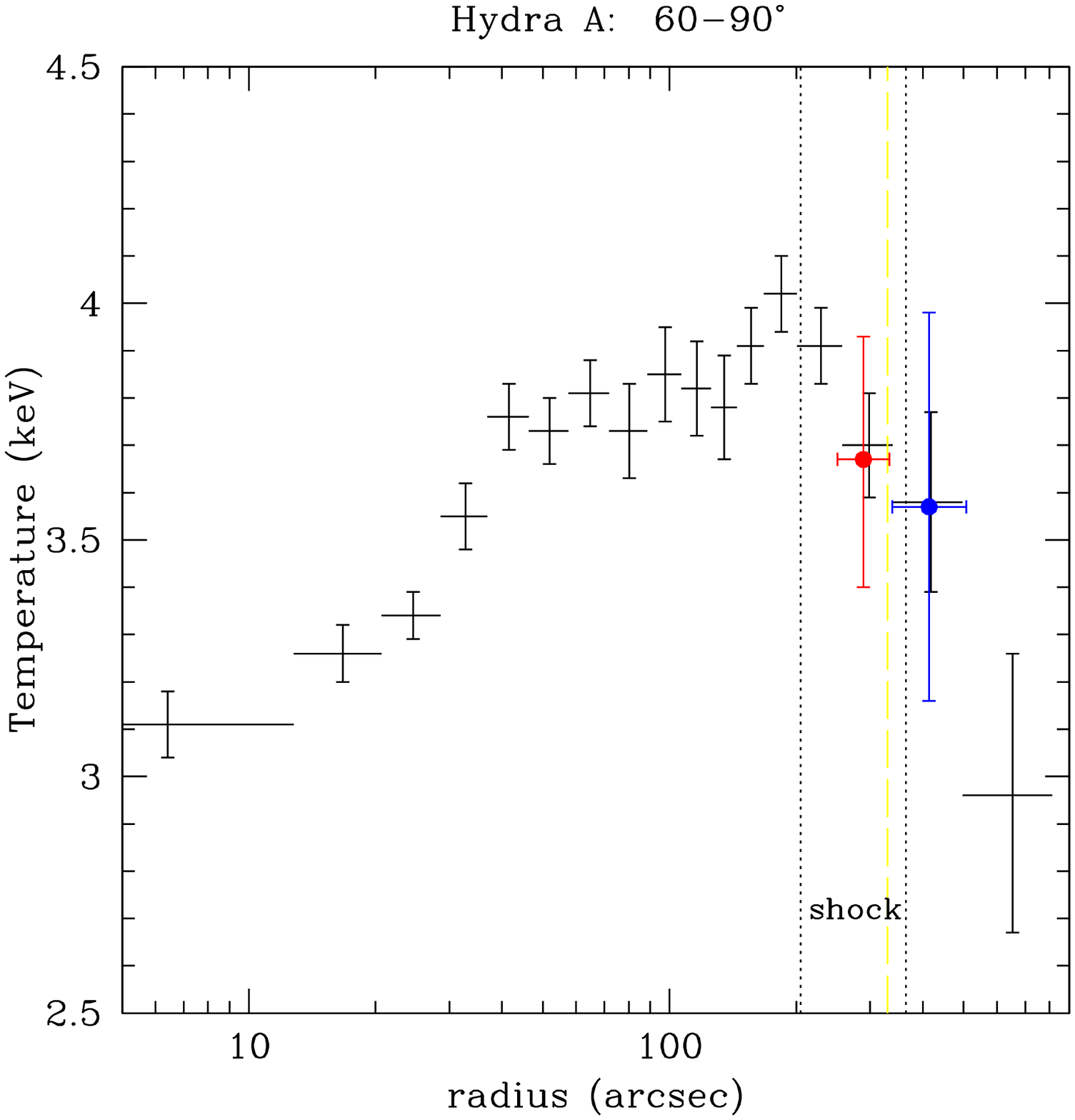}
}
\centerline{
\includegraphics[width=7.5cm]{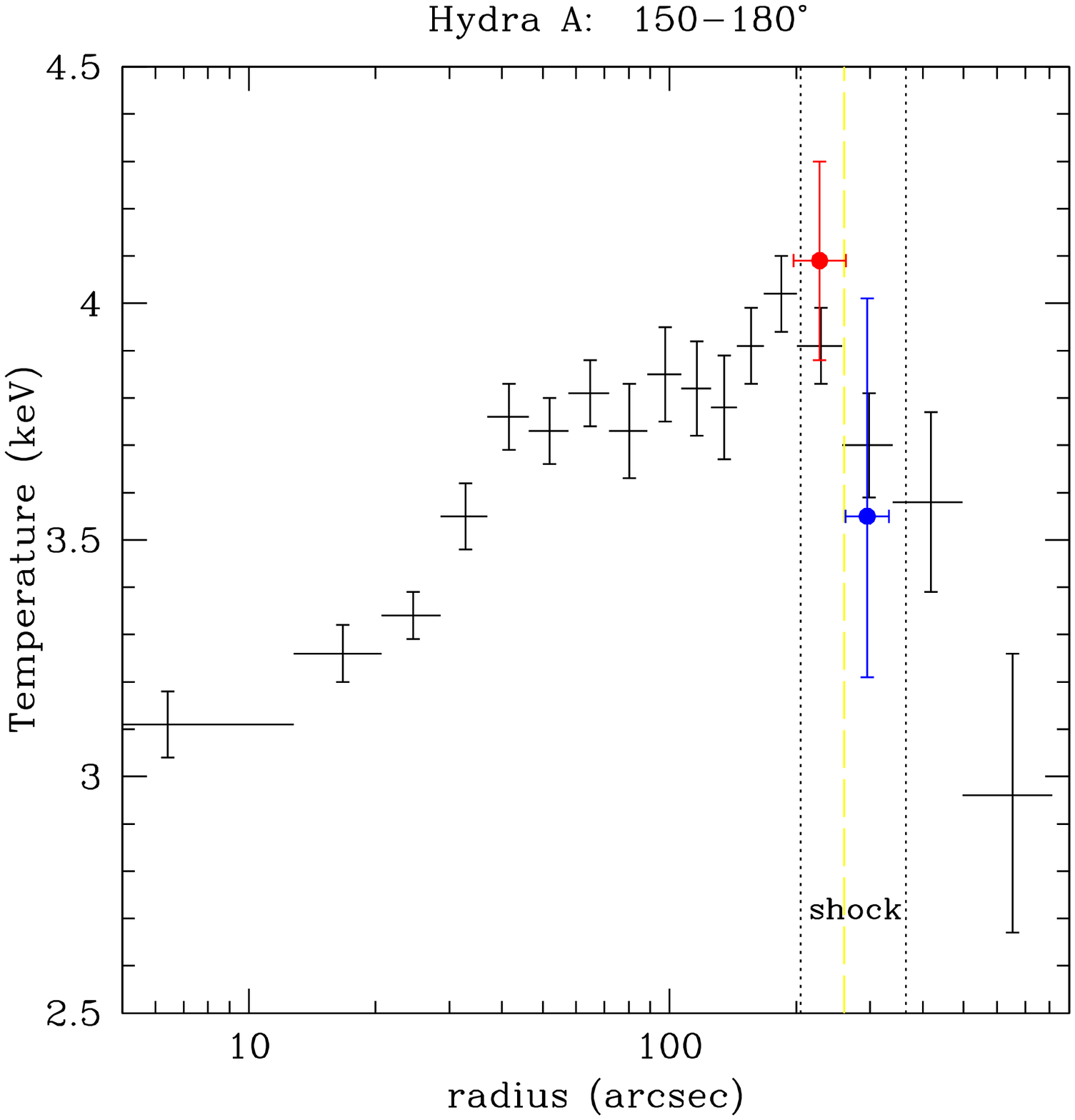}
\includegraphics[width=7.5cm]{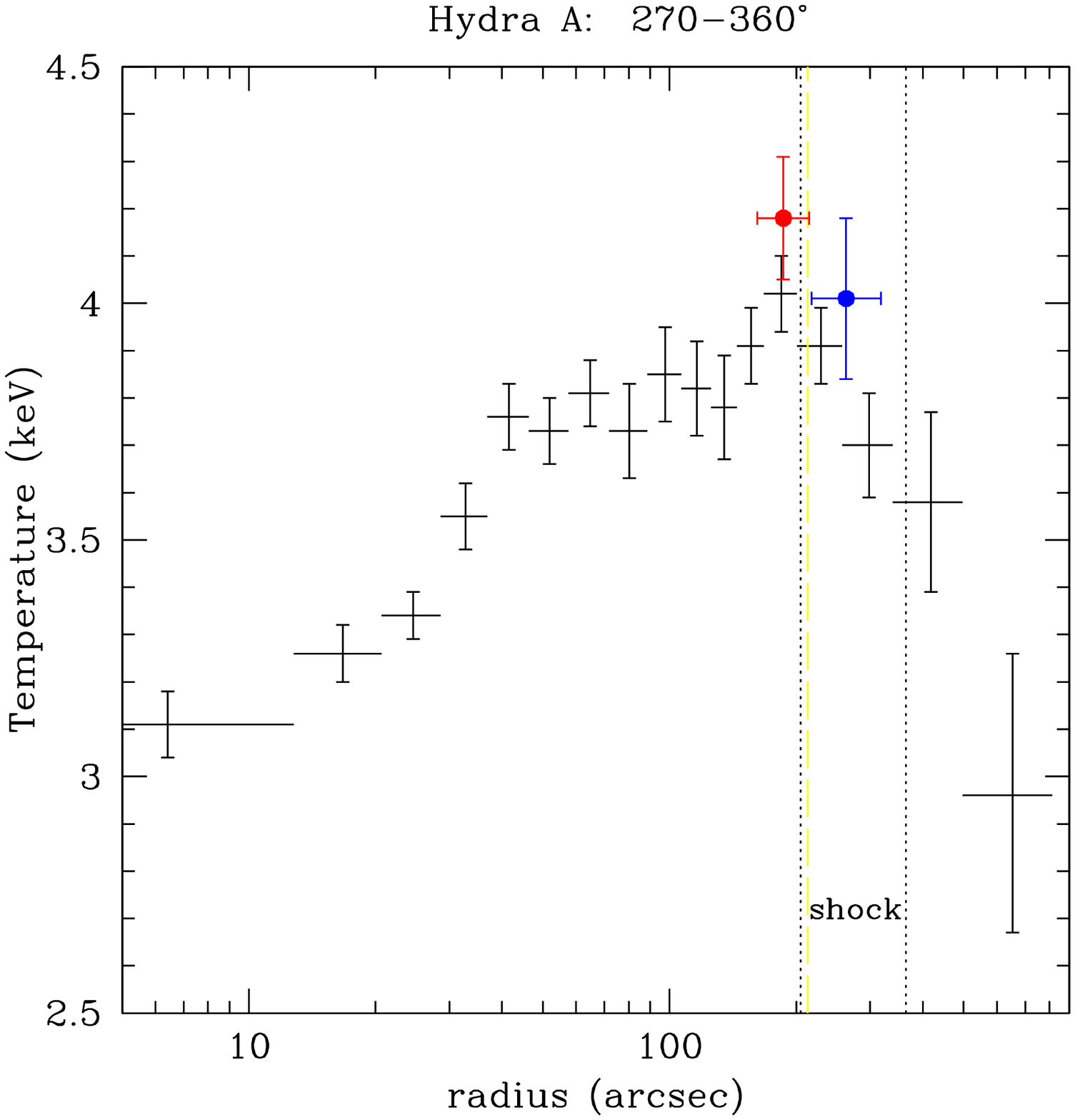}
}
\centerline{
\includegraphics[width=7.5cm]{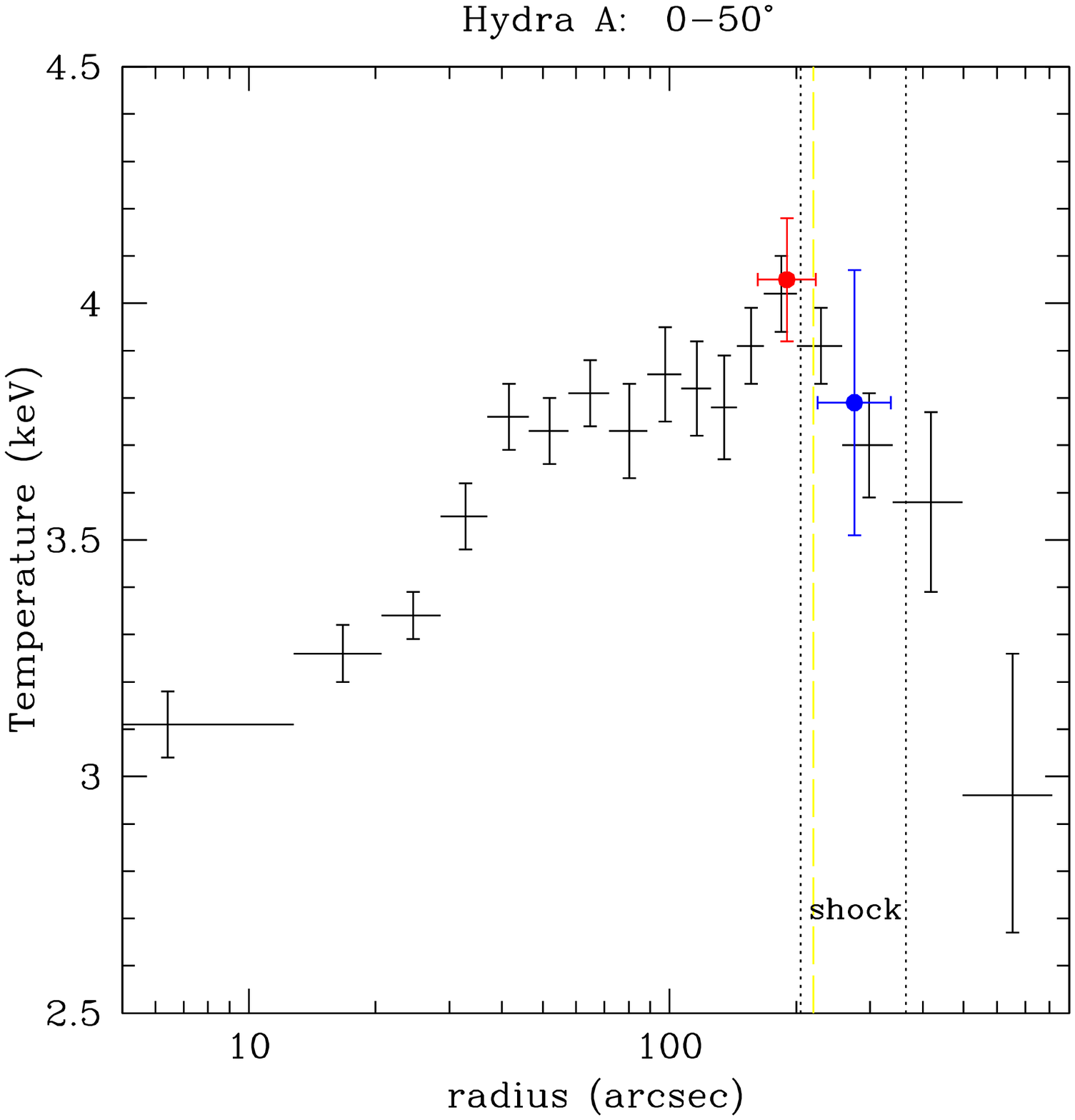}
\includegraphics[width=7.5cm]{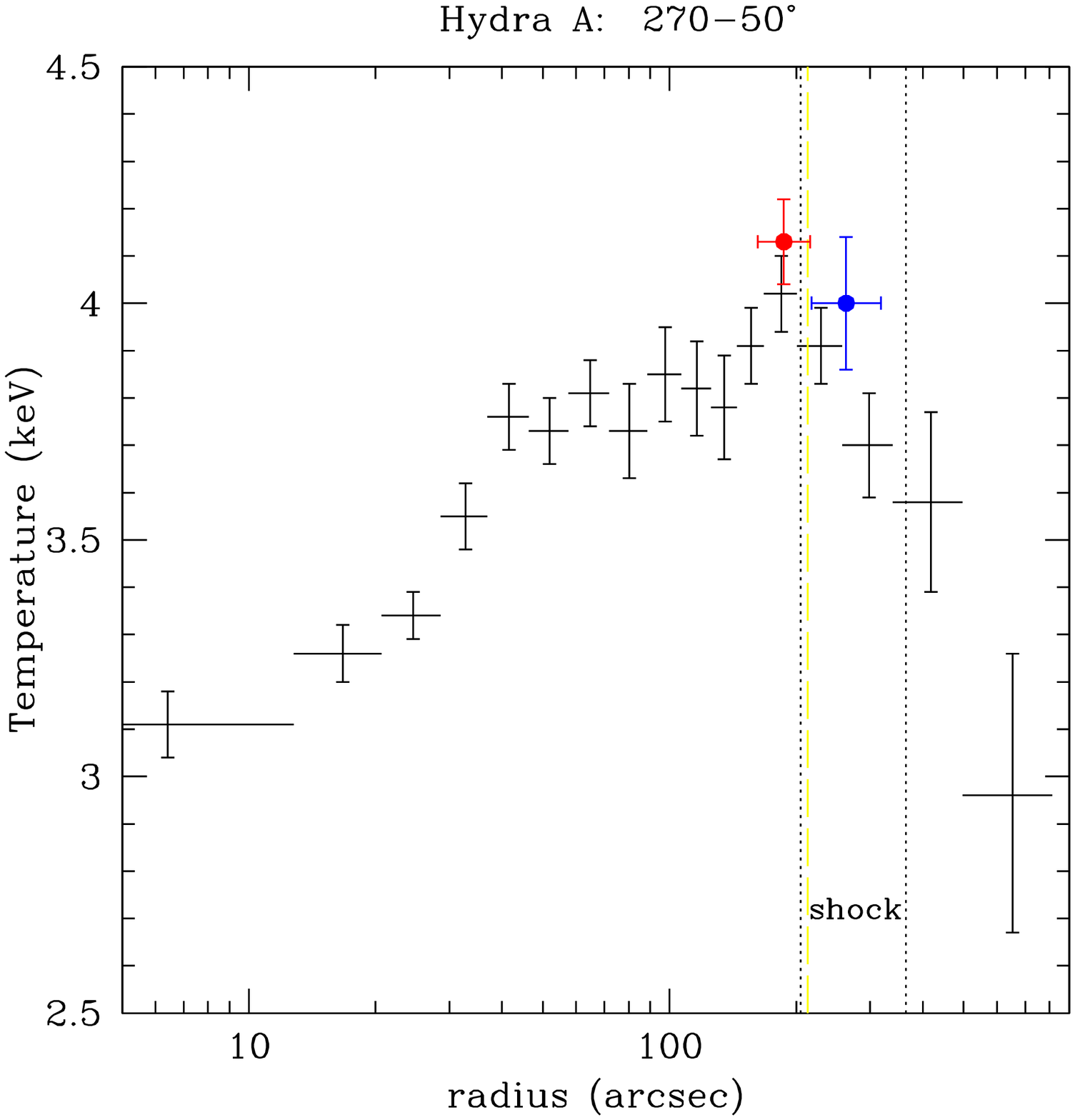}
}
\caption{\scriptsize \label{temp-shock.fig} Temperatures measured in
  the pre-shock (blue) and post-shock (red) regions of some of the
  sectors shown in Fig. \ref{shock_image.fig}. Radial error bars show
  the limits of the regions. The underlying global cluster profile,
  measured after masking the cool filaments corresponding to the
  profile shown in the right panel of Fig. \ref{temp.fig}, is shown in
  black.  The yellow dashed line indicates the radius of the shock
  front in each sector, while the dotted lines indicate the inner and
  outer radii of the shock front as determined in
  Sect. \ref{azimuthal.sec}.  The bottom panels are for binned
  sectors.  }
\end{figure*}


In fact, once we overlay the scaled temperature profile of Hydra A
measured after masking the cool filaments (right panel of
Fig. \ref{temp.fig}) it is remarkable how well it agrees with the
general shape of the temperature profiles observed for relaxed
clusters (see right panel of Fig. \ref{scaled_temp.fig}). Therefore,
it is unlikely that the temperature peak observed in Hydra A is
produced by the shock.  In particular, the spectroscopic detection of
a temperature rise in the regions immediately inside of the shock
front is complicated by the underlying rising temperature profile of
the global cluster atmosphere.  In Fig.  \ref{temp-shock.fig} we show
some examples of such temperature measurements in various sectors (see
Sect. \ref{temp-shock.sec}): the post-shock regions are found to be
hotter than the corresponding pre-shock ones, although due to the
large error bars, the pre-shock and post-shock temperatures are
consistent.


\subsection{Evidence for Gas Dredge-Up along the Cool Filaments}

The fact that the scaled temperature profile of Hydra A measured after
masking the filaments agrees with the general shape of the temperature
profiles observed for relaxed clusters (right panel of
Fig. \ref{scaled_temp.fig}) is a clear indication that these filaments
are responsible for the plateau of cool gas. In Section
\ref{spectral_cold.sec} we performed a detailed (projected) spectral
analysis and found evidence for multiphase gas, which may have been
uplifted with the radio lobes, giving rise to the cooler filaments
and plateau.  In order to estimate the mass of the cool gas in the
filaments, we must correct for the effects of projection along the
line-of-sight. Starting from the annuli used to derive the global
temperature profile, we binned them together in order to obtain three
shells suitable for a simple deprojection analysis of the cool
filaments. In particular, using three annuli: inner (0-72$'' \sim$0-76
kpc ), middle (72-145$'' \sim$76-152 kpc) and outer (145-339$''
\sim$152-360 kpc), we performed a deprojection analysis of the
absorbed 2T model with the XSPEC {\ttfamily projct} model. Under the
assumption of spherical shells of emission, this model calculates the
geometric weighting factors, according to which the emission is
redistributed amongst the projected annuli. The outer shell was
assumed to include only gas at the ambient temperature, which is
linked to the temperature values of the hot components in the inner
and middle shells.  The detailed results of the deprojection analysis
are shown in Table \ref{deproj.tab}.

We found that the hot phase has a temperature of 4 keV, in agreement
with the observed global temperature profile (Fig. \ref{temp.fig}),
and that the cool component is, on average, at $\sim$1.6 keV.  By
taking into account the relative filling factor, $f$, calculated under
the assumption that the two thermal components are in pressure
equilibrium in each shell, in the middle shell we estimated a mass of
cool gas of 9.9 $\times 10^{10} M_{\odot}$ (and a mass of hot gas of
1.7 $\times 10^{12} M_{\odot}$).  We note that this estimate is
considerably larger than the mass of cool gas reported by Simionescu
et al. (2009a). The discrepancy between the two measurements can be
ascribed to the fact that we included the mass content of the
mushroom-cap (namely, sectors ENE, NNE, NNW in Fig. \ref{cold.fig}).
Using the entropy of the cooler phase and comparing to the entropy
profile of the Hydra A cluster measured by David et al. (2001), we
could estimate where it originated.  From the density and temperature
values measured in the middle shell (see Table \ref{deproj.tab}), the
cool gas has an entropy $ S=kT_{\rm cool} \, n_{\rm e,cool}^{-2/3}
\sim$30 keV cm$^2$. As shown by the profile of David et al.  (2001),
gas at such entropy is located around 30 kpc from the cluster center.
Therefore, it is very likely that the cool gas, which is now observed
in the filaments at $\sim$75-150 kpc, was lifted from the central 30
kpc.  In this case the most obvious way to move the gas is by some
form of entrainment or dredge up of cool material from the center of
the cluster in the rising lobes. Churazov et al. (2000) have indeed
shown that hot buoyant bubbles produced by the expanding central radio
source can dredge up cool material from the center of the cluster.
The mass of cool gas at $\sim$75-150 kpc is 9.9 $\times 10^{10}
M_{\odot}$, which is $\sim$60\% of the total mass of gas remaining
within 30 kpc (1.7 $\times 10^{11} M_{\odot}$, David et al. 2001).
Assuming such a mass was lifted out of the central 30 kpc by a
continuous outflow or a series of bursts from the nucleus of the
central galaxy over the past 200-500 Myr (which created the cavity
system, Wise et al. 2007), it would amount to outflows of a few
hundred $M_{\odot}$ yr$^{-1}$. There would thus be a development of a
convectively unstable region that can significantly reduce the net
inflow of cooling gas (see also David et al.  2001, Nulsen et
al. 2002). Therefore our results show that the AGN feedback in Hydra A
is acting not only by directly re-heating the gas, but also by
removing a substantial amount of potential fuel for the supermassive
black hole (SMBH).

The energy required to lift the cool gas gives a lower limit on the
amount of AGN outburst energy deposited in the ICM.  This value can be
estimated by calculating the variation in gravitational potential
energy during the lifting process.  If we assume that the undisturbed
ICM is approximately isothermal with sound speed $c_{\rm s} \approx
1000$ km s$^{-1}$ and is in a hydrostatic configuration with density
profile $\rho (r)$, we can calculate this quantity as (Reynolds et
al. 2008)
\begin{equation}
  \Delta E = \frac{M_{\rm cool} \, c_{\rm s}^2}{\gamma} \ln 
             \left( \frac{\rho_i}{\rho_f} \right)
\end{equation}
where $M_{\rm cool}$ is the lifted mass, $\rho_i$ and $\rho_f$ are the
initial and final densities of the surrounding ICM, and $\gamma$=5/3
is the ratio of specific heat capacities. From the density profile
presented by David et al. (2001) we estimated that the energy required
to lift the cool gas is \gtsim $2.2 \times 10^{60}$ erg.  This value
is comparable to the work required to inflate all of the cavities
against the surrounding pressure (Wise et al. 2007) and is $\sim$25\%
of the total energy of the large-scale shock (Nulsen et al. 2005).
Although we find evidence for a more extended gas dredge-up than
previously estimated, there is a remarkable correlation between the
cool filaments studied here and the metal-rich filaments in the
iron-abundance maps measured by Simionescu et al. (2009a) and
Kirkpatrick et al. (2009). This is consistent with the idea that Hydra
A's powerful radio source is able to lift cool, metal-rich gas from
the central region and distribute it throughout the X-ray atmosphere
of the cluster.  A similar effect is observed in M87 (Simionescu et al
2008, Werner et al. 2010).  We finally note that by summing our
estimates of the integrated mass of cool gas, $M_{\rm cool} (<152 {\rm
  \, kpc}) \sim$1.7 $\times 10^{11} M_{\odot}$, and hot gas, $M_{\rm
  hot} (<152 {\rm \, kpc}) \sim$2.3 $\times 10^{12} M_{\odot}$, in the
inner and middle shells we measure a total mass of gas $M_{\rm gas}
(<152 {\rm \, kpc}) \sim$2.5 $\times 10^{12} M_{\odot}$ which is in
agreement with the gas mass profile measured by David et al. (2001),
so our general picture is self-consistent.


\section{Summary}

The main results of this work can be summarized as follows:

\begin{itemize}

\item We studied the azimuthal properties of the weak (Mach number
  $\sim$1.2-1.3), large-scale shock and attempted to measure the
  temperature jump associated with the shock in different directions.
  The errors in the temperature measurements are too large to
  constrain the temperature jump caused by the shock. Furthermore, we
  note that the detection of a temperature rise in the regions
  immediately inside of the front is complicated by the underlying
  rising temperature profile of the cluster atmosphere.

\item Our detailed spectral analysis indicates the presence of
  multiphase gas along soft filaments seen in the hardness ratio map.
  The cooler gas has a significant impact on the radial temperature
  profile of the cluster.  After correcting for the effect of the cool
  filaments, Hydra A's temperature profile is consistent with the form
  observed in relaxed galaxy clusters.  Thus it is unlikely that the
  observed temperature peak is produced by the shock.

\item The cool filaments are direct evidence for substantial dredge-up
  of low entropy material by the rising lobes: $\sim$60\% of the gas
  mass remaining in the central 30 kpc has been lifted along the cool
  filaments to the observed current position of 75-150 kpc. The
  outflow amounts to a few hundred $M_{\odot}$ yr$^{-1}$, which is
  comparable to the inflow rate formerly estimated for the cooling
  flow. The energy required to lift the cool gas is $\sim$25\% of the
  total energy of the large-scale shock and is comparable to the work
  required to inflate the cavities.

\end{itemize}

\acknowledgments

We thank the referee for constructive comments.  We thank A. Vikhlinin
for providing the data used to make the plots in
Fig. \ref{scaled_temp.fig}.  MG thanks D. Rafferty for precious advice
in data reduction, and F. Brighenti for many insightful
discussions. MG acknowledges support by grants ASI-INAF I/023/05/0 and
I/088/06/0 and by Chandra grants GO0-11003X and GO0-11136X.  BRM
acknowledges support from the Natural Sciences and Engineering
Research Council of Canada.


\end{document}